\documentclass[12pt]{custarticle}

\usepackage{hyperref}
\usepackage{float}
\usepackage{verbatim} 
\usepackage{apalike}
\usepackage{mathtools} 
\usepackage{booktabs} 
\usepackage{tikz} 
\usepackage{color}
\usepackage{bbm}
\usepackage[frozencache,cachedir=minted-cache]{minted}
\usepackage{minted}
\usepackage{geometry}
\usepackage{setspace}
\usepackage{graphicx}	
\usepackage{algorithm}
\usepackage{multicol}
\usepackage{amsthm}
\usepackage{amssymb}
\usepackage{mathtools}
\usepackage{subcaption}
\usepackage{tcolorbox}

\graphicspath{ {./images/} }

\newcommand{\rev}{R}
\makeatletter
\def\ps@pprintTitle{%
  \let\@oddhead\@empty
  \let\@evenhead\@empty
  \let\@oddfoot\@empty
  \let\@evenfoot\@oddfoot
}
\makeatother

\newcommand{\name}{\textit{InteraSSort}}
\newcommand{\rec}{\textit{InteraRec}}
 \biboptions{authoryear}

\begin{document}
\begin{frontmatter}

\title{\rec{}: Screenshot Based Recommendations Using Multimodal Large Language Models}

\author[label1]{Saketh Reddy Karra\corref{cor1}}
\ead{skarra7@uic.edu}

\author[label1]{Theja Tulabandhula}
\ead{theja@uic.edu}

\address[label1]{University of Illinois Chicago, 601 S Morgan St, Chicago, IL 60607, United States}

\begin{abstract}
Weblogs, comprised of records detailing user activities on any website, offer valuable insights into user preferences, behavior, and interests. Numerous recommendation algorithms, employing strategies such as collaborative filtering, content-based filtering, and hybrid methods, leverage the data mined through these weblogs to provide personalized recommendations to users. Despite the abundance of information available in these weblogs, identifying and extracting pertinent information and key features from them necessitate extensive engineering endeavors. The intricate nature of the data also poses a challenge for interpretation, especially for non-experts. In this study, we introduce a sophisticated and interactive recommendation framework denoted as \rec{}, which diverges from conventional approaches that exclusively depend on weblogs for recommendation generation. \rec{} framework captures high-frequency screenshots of web pages as users navigate through a website. Leveraging state-of-the-art multimodal large language models (MLLMs), it extracts valuable insights into user preferences from these screenshots by generating a textual summary based on predefined keywords. Subsequently, an LLM-integrated optimization setup utilizes this summary to generate tailored recommendations. Furthermore, we explore the integration of session-based recommendation systems into the \rec{} framework, aiming to enhance its overall performance. Finally, we curate a new dataset comprising of screenshots from product web pages on the Amazon website for the validation of the \rec{} framework. Detailed experiments demonstrate the efficacy of the \rec{} framework in delivering valuable and personalized recommendations tailored to individual user preferences.
\end{abstract}


\end{frontmatter}

\section{Introduction} \label{sec:introduction}
In the evolving landscape of e-commerce, the need to understand user preferences for offering personalized content is increasingly crucial for online platforms to maximize revenue, build customer loyalty, and maintain competitive advantage. Typically, the nuanced information that can explain the user browsing behavior on these platforms is stored as weblogs~\citep{rosenstein2000actually}. Numerous state-of-the-art recommendation systems, employing strategies such as collaborative filtering, content-based filtering, and hybrid methods, leverage the data mined from weblogs for both training and evaluation tasks. This data-driven approach ensures that recommendations align closely with user preferences, contributing to a more seamless and satisfying shopping journey for users and ultimately increasing overall platform revenue.

Raw weblogs contain a wealth of information on key browsing session details, including session start and end times, visited pages, and click-stream data. Beyond capturing fundamental details, these weblogs delve into user-specific data, encompassing identifiers and IP addresses. However, interpreting raw weblog data directly can be challenging for non-experts, requiring them to navigate its complexity and processing it to build recommendation models. Often, sophisticated data engineering techniques are necessary to extract the relevant features and datasets required for training these models. 

Addressing the above issues with processing weblogs, we propose an innovative approach leveraging the screenshots of users' browsing activities. By opting for screenshots instead of relying on weblogs, the system can benefit from heightened interpretability. Notably, the visual nature of screenshots offer a lucid and transparent representation of user actions, thus significantly enhancing the ability to draw meaningful insights. In contrast to processing weblogs, which can introduce complexities associated with extracting various features, screenshots simplify the inputs, making it a more straightforward and intuitive process. Additionally, harnessing screenshots instead of weblogs facilitates real-time personalization, benefiting from the near-instantaneous capture of screenshots.

\begin{figure}[htbp]
  \centering
  \includegraphics[width=1\linewidth]{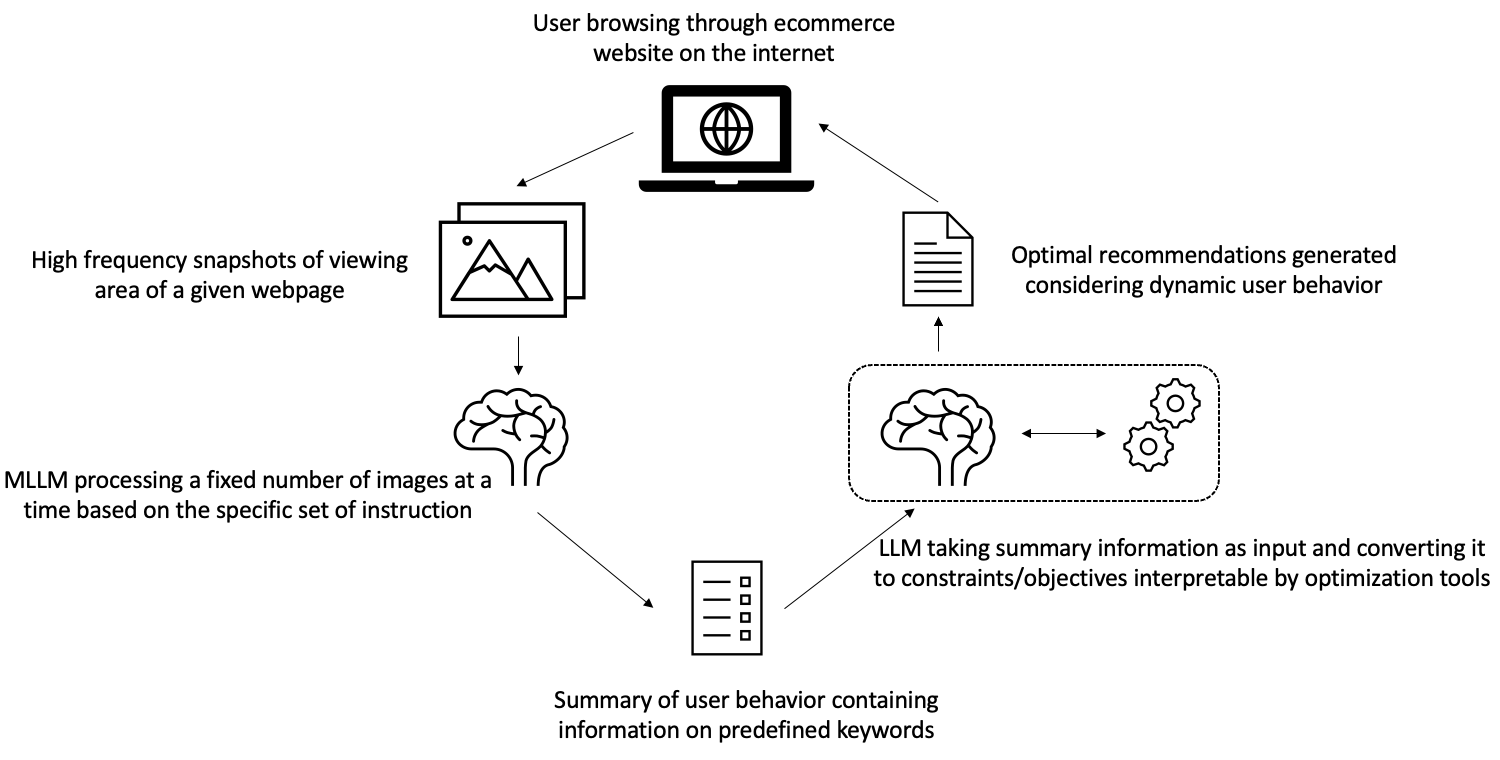}
  \caption{Schematic representation of the \rec{} framework.}
  \label{fig:frame}
\end{figure}

Utilizing screenshots to generate personalized recommendations necessitates additional processing. With recent advancements in the domain of LLMs, particularly MLLMs have shown impressive capabilities in performing complex tasks ranging from open world image understanding to interpreting memes~\citep{yang2023mm}. Due to their strong reasoning and in-context learning abilities, numerous researchers have been exploring their applications in both academia and industry~\citep{hadi2023survey}. This provides a compelling case for developing a robust framework using LLMs to process screenshots, presenting a superior alternative to relying solely on weblogs.

Inspired by the above, we introduce \rec{}, a novel screenshot-based recommendation framework as shown in the figure~\ref{fig:frame}. \rec{} initiates by capturing high-frequency screenshots of web pages as users navigate them in real-time. Subsequently, an MLLM processes these screenshots to extract meaningful insights into user behavior. These insights are then utilized in an LLM-integrated optimization setup to generate relevant and personalized recommendations. By seamlessly integrating visual data, language models, and optimization tools, \rec{} surpasses the limitations of existing systems, offering a more personalized and effective recommendation experience for users.

The \rec{} framework summarizes the extracted insights based on a predefined set of keywords, shedding light on browsing patterns and purchase intentions of users. This summary can be utilized to improve the performance of the existing recommendation systems. Therefore, we explore a special case of integrating the session-based recommendation models in the \rec{} framework. In particular, we leverage the keyword summaries extracted from screenshots of product web pages to re-rank the model outputs, to enhance the prediction accuracy. To evaluate the session-based recommendation models under the above setting, we manually curate a new dataset of screenshots from users browsing Amazon website. Our experiments findings validate the effectiveness of the proposed \rec{} framework particularly in scenarios with limited data availability.   

We summarize main contributions of our work below.
\begin{itemize}
    \item To the best of our knowledge, we are the first to use screenshots to generate personalized recommendations on online e-commerce platforms, moving away from the traditional weblog based approaches.
    \item We present \rec{}, an innovative recommendation framework that leverages screenshots as its core input. Harnessing the capabilities of both LLM and MLLMs, \rec{} extracts insights from the screenshots and intelligently executes suitable optimization tools, translating user behavior into concise, easily interpretable user recommendations.
    \item We leverage the keyword summaries generated using the \rec{} framework to improve the performance of existing session-based recommender models by re-ranking their predictions.
    \item We validate the effectiveness of integrating \rec{} with session-based recommendation models on a new dataset comprising of screenshots from user browsing sessions curated from the Amazon website.
\end{itemize}

The paper is structured as follows. We delve into closely related work in Section \ref{sec:related_work}. In Section \ref{sec:framework}, we discuss the \rec{} framework in three-stages and we delve into integrating the \rec{} framework with existing state-of-the-art session-based recommendation systems. In Section \ref{sec:experiments}, we present the illustrative examples of \rec{} framework and discuss the experiments to validate the performance of session-based recommendation models with \rec{} framework. Finally, we conclude in Section \ref{sec:conclusion} with some comments on future work.

\section{Related work} \label{sec:related_work}
In this study, we expand upon two key streams of research: (a) LLMs for user behavior modeling, and (b) Tool integration with LLMs. We briefly discuss some of the related works below.

\subsection{LLMs for user behavior modeling}
LLMs are known for their capability to dynamically generate diverse facets of user profiles by analyzing their historical viewing and transaction data.~\citep{chen2023palr} generated user profiles from television viewing history and and then utilized these profiles to retrieve candidate items from a pool, subsequently employing a LLM for item recommendations.~\citep{liu2023first} employed LLMs to create user profiles, exploring topics and regions of interest based on user browsing history, and integrated the inferred user profile to enhance recommendations.~\citep{zheng2024harnessing} utilized an LLM based summarizer to create user profiles from text rich item attributes for generating personalized recommendations. Unlike the approaches discussed above that rely on text data directly from user history for recommendation generation, our novel proposal creates dynamic user behavior summary by leveraging screenshots of user interactions using MLLMs.

\subsection{Tool integration with LLMs}
Researchers have made significant strides in using LLMs to tackle complex tasks by extending their capabilities to include planning and API selection for tool utilization. For instance,~\cite{schick2023toolformer} introduced the pioneering work of incorporating external API tags into text sequences, enabling LLMs to access external tools. TaskMatrix.AI \cite{liang2023taskmatrix} utilizes LLMs to generate high-level solution outlines tailored to specific tasks, matching subtasks with suitable off-the-shelf models or systems. HuggingGPT~\citep{shen2023hugginggpt} harnesses LLMs to connect to various external state-of-the-art transformer models for solving complex AI tasks effectively. Lastly,~\cite{qin2023tool} proposed a tool-augmented LLM framework that dynamically adjusts execution plans, empowering LLMs to complete subtasks using appropriate tools proficiently.~\cite{li2023large} introduced the Optiguide framework, leveraging LLMs to elucidate supply chain optimization solutions and address what-if scenarios. In contrast to the aforementioned approaches, \rec{} harnesses the power of LLMs to generate item recommendations by converting user behavior summaries into a format that can be readily interpreted by an optimization solver.

\section{The \rec{} framework} \label{sec:framework}
Addressing the challenge of offering real-time recommendations to users is a multi-step process. It typically begins with thorough data collection and analysis, followed by careful selection and training of appropriate user behavior models. After a rigorous evaluation, the trained model is deployed to generate optimal recommendations. The user is then empowered to make a decision, choosing whether to act upon or disregard the presented recommendations.

\begin{figure}[htbp]
  \centering
  \includegraphics[width=1\linewidth]{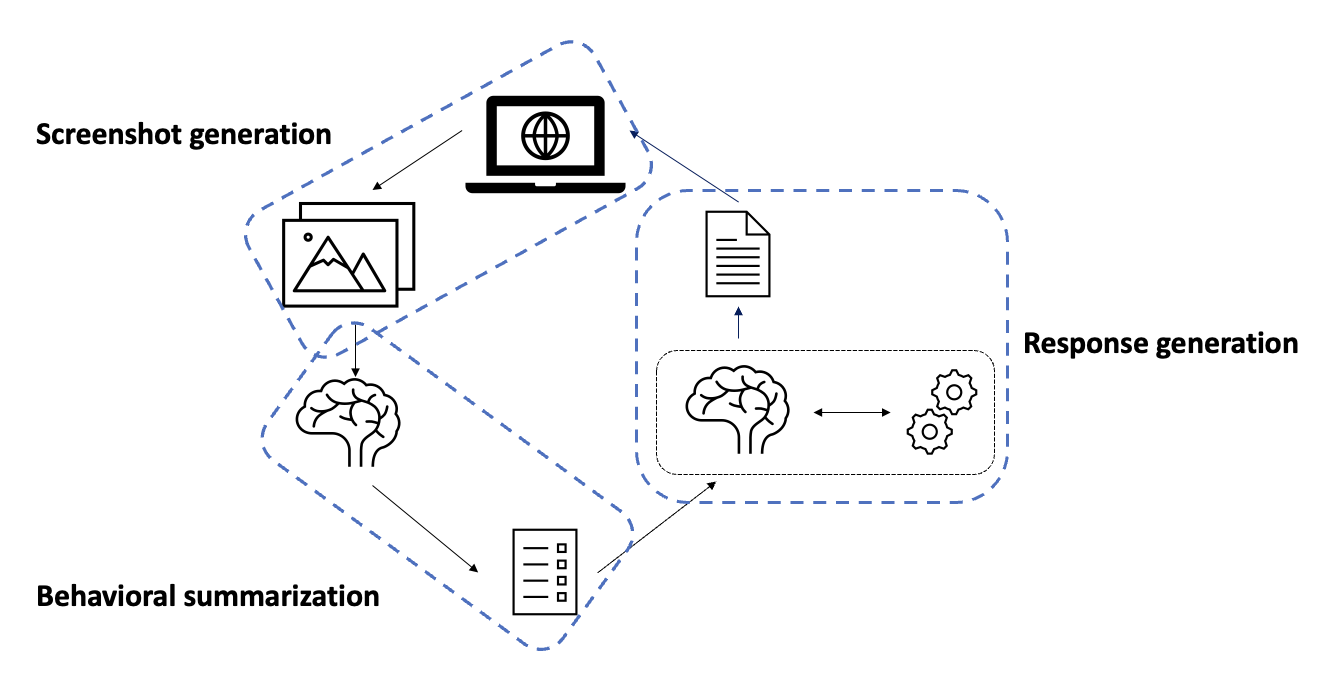}
  \caption{Overview of three stages of \rec{} framework.}
  \label{fig:chat}
\end{figure}

Our framework \rec{} initiates by capturing screenshots of a user's browsing activity on a website at regular intervals. Subsequently, an MLLM, with its inferential capabilities, translates user interaction behavior from screenshots to a text summary comprising of information on predefined keywords.  The \rec{} framework then leverages a suitable recommendation system to extract user preferences based on this summary and generate tailored recommendations. The process outlined above is organized into multiple stages:  1) Screenshot generation, 2) Behavioral summarization, and 3) Response generation as shown in Figure~\ref{fig:chat}.

\subsection{Screenshot generation}
\rec{} initiates an automated script to systematically capture high-frequency, high-quality screenshots of the web pages users navigate during a browsing session. Notably, the screenshots are confined to the user's current viewing area on the screen rather than encompassing the full scrollable content of each webpage. This allows for efficient targeting of the visible content that users actively engage with. \rec{} then stores these representative screenshots in a database for further processing and analysis.
\begin{figure}[htbp]
\begin{tcolorbox}[colframe=cyan, colback=white!10]
    \begin{minted}{python}
response = client.chat.completions.create(
model="gpt-4-vision-preview"
messages=[
{
"role": "user"
"content": [
{
"type": "text"
"text": "What can you infer from the images below \
with regards to a user preference in the following categories? \
Product Characteristics, \
Lowest Price, \ 
Highest Price, \ 
Brand Preference, \ 
Product Specifications, \ 
User Reviews and Testimonials, \ 
Comparisons, \
Promotions. \
Write a response that contains the above information. \ 
if any of the categorical information is unavailable, \
mark it as not available "}...],...}
\end{minted}
\end{tcolorbox}
\caption{Guidelines for MLLM to generate a summary of user interactions using predefined keywords}
\label{fig:keywords}
\end{figure}

\subsection{Behavioral summarization}
In this stage, \rec{} sequentially processes a finite number of screenshots in real-time using an MLLM to analyze and provide detailed responses. Specifically, \rec{} instructs the MLLM to succinctly summarize user interaction behavior across predefined categories such as product characteristics, lowest and highest prices, brand preferences, product specifications, user reviews, comparisons, and promotions as shown in Figure~\ref{fig:keywords}. This precise instruction is crucial to filter the information and capture elements directly aligned with user interests. Moreover, generic instructions like ``describe the user behavior on screenshots" or ``explain the differences in screenshots"  might result in responses lacking specificity, thereby failing to capture user preferences. Finally, \rec{} ensures that the model presents information in a summarized JSON format, facilitating its use as a constraint or filter for subsequent processing.

\begin{figure}[htbp]
\begin{tcolorbox}[colframe=cyan, colback=white!10]
    \begin{minted}{python}
function_descriptions = [
    {
    "name": "get_user_recommendations", 
    "description": "Generate dynamic  recommendations based \
                    the summary of user behavior", 
    "parameters": {
    "type": "object", "properties": {
        "lowest_price": { 
            "type": "integer", 
            "description": "get lowest price preference of user."
        ,},
        "highest_price": {
             "type": "integer", 
             "description": "get highest price preference of user."
        ,},
        "color": {
            "type": "string",
            "description": "get color preference of user",
        }, }, }, }, ]
\end{minted}
\end{tcolorbox}
\caption{Prospective function for decomposing user interaction summaries.}
\label{fig:function}
\end{figure}

\subsection{Response generation}
The keyword-based summary generated from the previous stage is rich in information capturing user preferences, forming a valuable resource for generating recommendations. The relevant information to be extracted from the summary depends on the capabilities of recommendation methods that can incorporate them. In this particular setting, \rec{} harnesses the \name{} framework~\citep{karra2023interassort} to deconstruct the user behavior summary into relevant constraints and uses them to solve an assortment optimization problem to generate optimal recommendations. Specifically, \rec{} leverages the function-calling capabilities of the LLM to decompose the user interaction summary into relevant constraints using appropriate functions as shown in Figure~\ref{fig:function}. Subsequently, rigorous validation checks are conducted, encompassing range and consistency assessments of the decomposed constraints. \rec{} maintains an extensive database containing parameters for discrete choice models estimated using historical purchase transactions. By leveraging these choice model parameters and additional constraints as arguments, \rec{} executes optimization scripts employing tools like optimization solvers to generate optimal solutions. Ultimately, \rec{} empowers the LLM to incorporate these results as input, generating user-friendly personalized product recommendations.

\subsection{Special case of response generation with session-based recommendation model} 
\begin{figure}[htbp]
  \centering
  \includegraphics[width=1\linewidth]{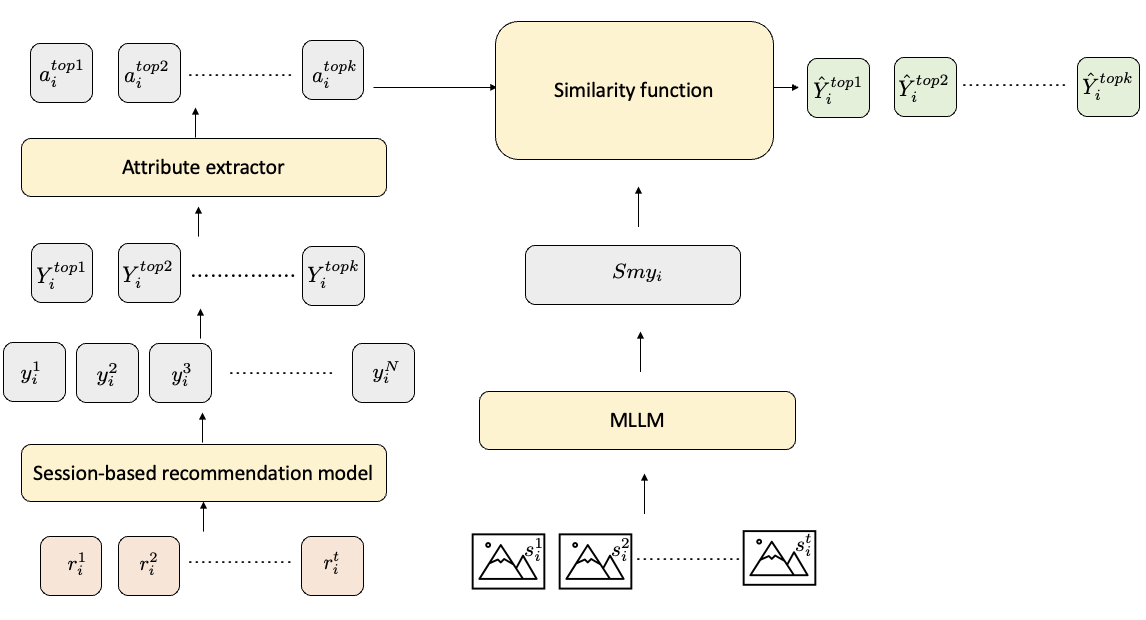}
  \caption{Session-based recommendations with \rec{} framework}
  \label{fig:session_based_figure}
\end{figure}

The response generation stage of the \rec{} framework can be seamlessly integrated with existing recommendation systems to enhance their overall performance and effectiveness. Specifically, we explore the integration of session-based recommender models within the \rec{} framework. The schematic diagram in Figure~\ref{fig:session_based_figure} illustrates the proposed setup, which combines the strengths of the \rec{} framework and session-based recommender models to deliver superior recommendations.

Session-based recommendation aims to predict the items a user is most likely to engage with next, given their chronological sequence of interactions during a browsing session. Let $R_i = [r_i^1,\dots,r_i^t]$ represent the sequence of $t$ items the user has interacted with (e.g., clicked, purchased) in session $i$. A session-based recommendation model takes $R_i$ as input and outputs $Y_i = [y_i^1,\dots, y_i^N]$, where $y_i^j$ is the prediction probability of the user interacting with item $j$ in the subsequent time step. The model's performance is then evaluated by comparing the top-k ranked items in $Y_i$, denoted as $[\hat{y}_i^{top1},\dots,\hat{y}_i^{topk}]$, against the ground truth data.

Given the top-k ranked items, we employ a re-ranking strategy that incorporates the user's browsing context, derived from screenshots of web pages viewed during their session, using the \rec{} framework. This approach re-ranks the initial top ranked predictions from the session-based recommendation model, leveraging the contextual information extracted from the screenshots to better align the recommendations with the user's intent and browsing behavior. Specifically, we collect the screenshots of item web pages corresponding to the user's past interactions in the current session, denoted as $S_i = [s_i^1,\dots, s_i^t]$, where $s_i^j$ represents the screenshot of the $j$-th item interacted with. The MLLM component of \rec{} processes these screenshots to generate a keyword summary $Smy_i$, capturing the user's interests and preferences during the session. Additionally, each item has associated attributes such as color, brand, price, title, etc. For the top-k ranked item predictions, we extract the concatenated textual attributes, denoted as $[a_i^{top1},\dots, a_i^{topk}]$, where $a_i^{topj}$ represents the attributes of the $j$-th ranked item.

We leverage a sentence embedding model to obtain vector representations for both the keyword summary and the concatenated text attributes. By computing the cosine similarity between the summary embeddings and the attribute embeddings of each top-ranked item, we re-rank the initial predictions in descending order of their similarity scores. This re-ranking process yields the final ranked list $[\hat{y}_i^{top1},\dots, \hat{y}_i^{topk}]$. By aligning the recommendations with the user's interests and preferences, captured by the keyword summary derived from their browsing session, we ensure that the final recommendations are highly relevant and personalized to the users.

\section{Experiments} \label{sec:experiments}

\subsection{Example demonstration of \rec{} framework}
In this example, we configure the \rec{} framework to solve the assortment planning problem under a discrete choice model to generate the personalized recommendations as discussed below.

\subsubsection{Assortment planning}
The assortment planning problem involves choosing an assortment among a set of feasible assortments ($\mathcal{S}$) that maximizes the expected revenue. Consider a set of items indexed from $1$ to $n$ with their respective prices being $p_1, p_2, \cdots p_n$. The revenue of the assortment is given by $\rev(S) = \sum_{k \in S} p_k \times \mathbb{P}(k|S)$ where $S \subseteq \{1,...,n\}$. The expected revenue maximization problem is simply: $\max_{S \in \mathcal{S}} \rev(S)$. Here $\mathbb{P}(k|S)$ represents the probability that a user chooses item $k$ from an assortment $S$ and is determined by a choice model. 

The complex nature of the assortment planning problem requires the development of robust optimization methodologies that can work well with different types of constraints and produce viable solutions within reasonable time frames. In our experiments, we adopt a series of scalable efficient algorithms~\citep{tulabandhula2022optimizing} for solving the assortment optimization problem.

\subsubsection{Multinomial logit (MNL)}
The MNL model is one of the most extensively studied discrete choice models and is frequently utilized across various marketing applications. The parameters of the MNL model are represented by a vector $\mathbf{v} = \left(v_0, v_1, \cdots v_{n}\right)$ with $0 \leq v_i \leq 1 \;\;\forall i$. Parameter $v_i, \ 1\leq i \leq n $, captures the user preference for purchasing item $i$. Under this model, the probability that a user chooses item $k$ from an assortment $S$ is given by $\mathbb{P}(k|S) = v_l/(v_0 + \sum_{k' \in S} v_{k'})$.

\subsubsection{Model settings}
We employ the \texttt{gpt-4-vision-preview} (GPT-4V) and \texttt{gpt-3.5-turbo} variants from the GPT model series as our MLLM for processing screenshots and LLM for decomposing the keyword summary respectively. Both the models are publicly accessible through the OpenAI API~\footnote{\url{https://platform.openai.com/}}.

\subsubsection{Illustrative example}
\begin{figure}[htbp]
  \begin{multicols}{3}
    \includegraphics[width=\linewidth]{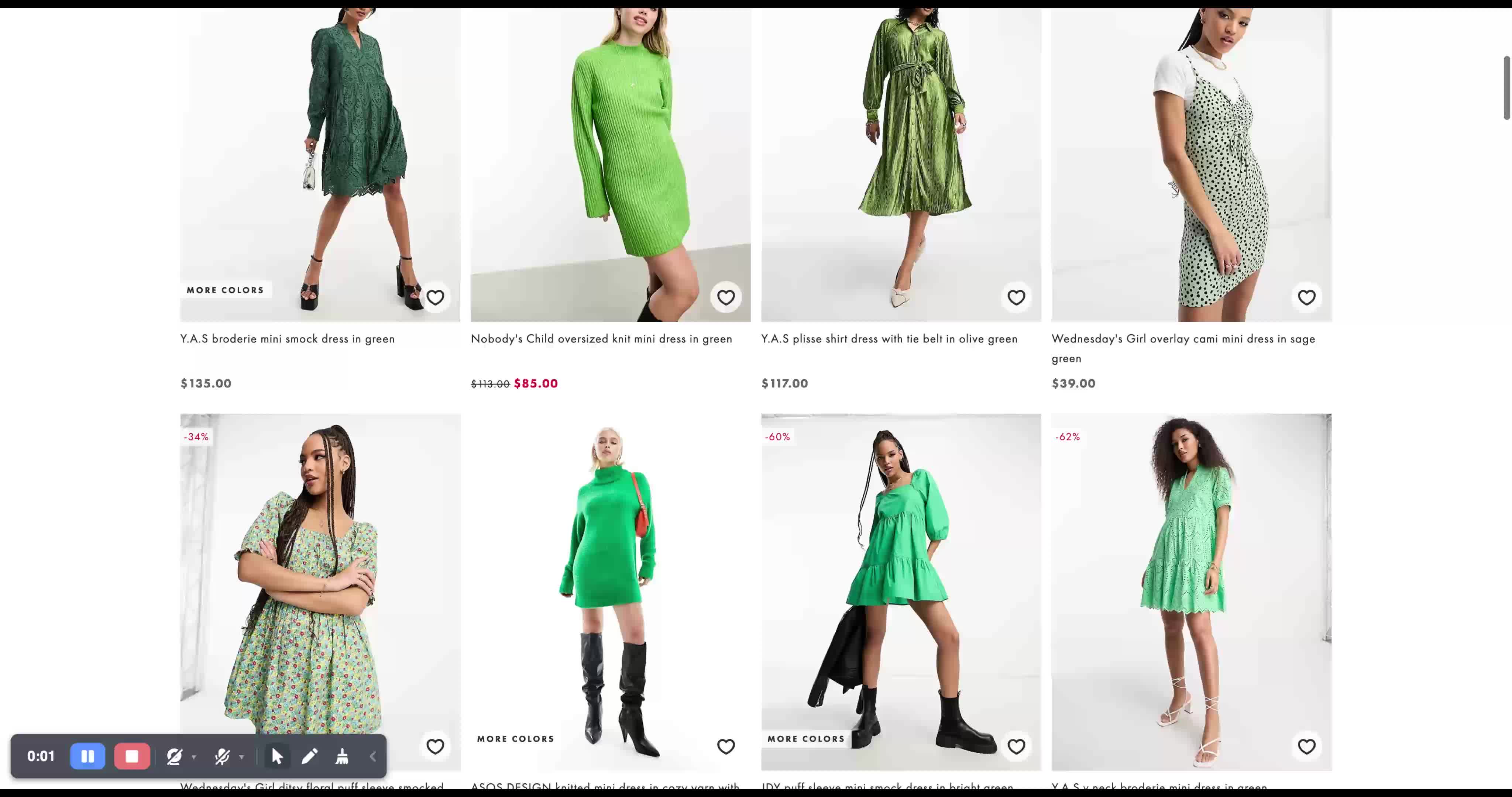}\par
    \includegraphics[width=\linewidth]{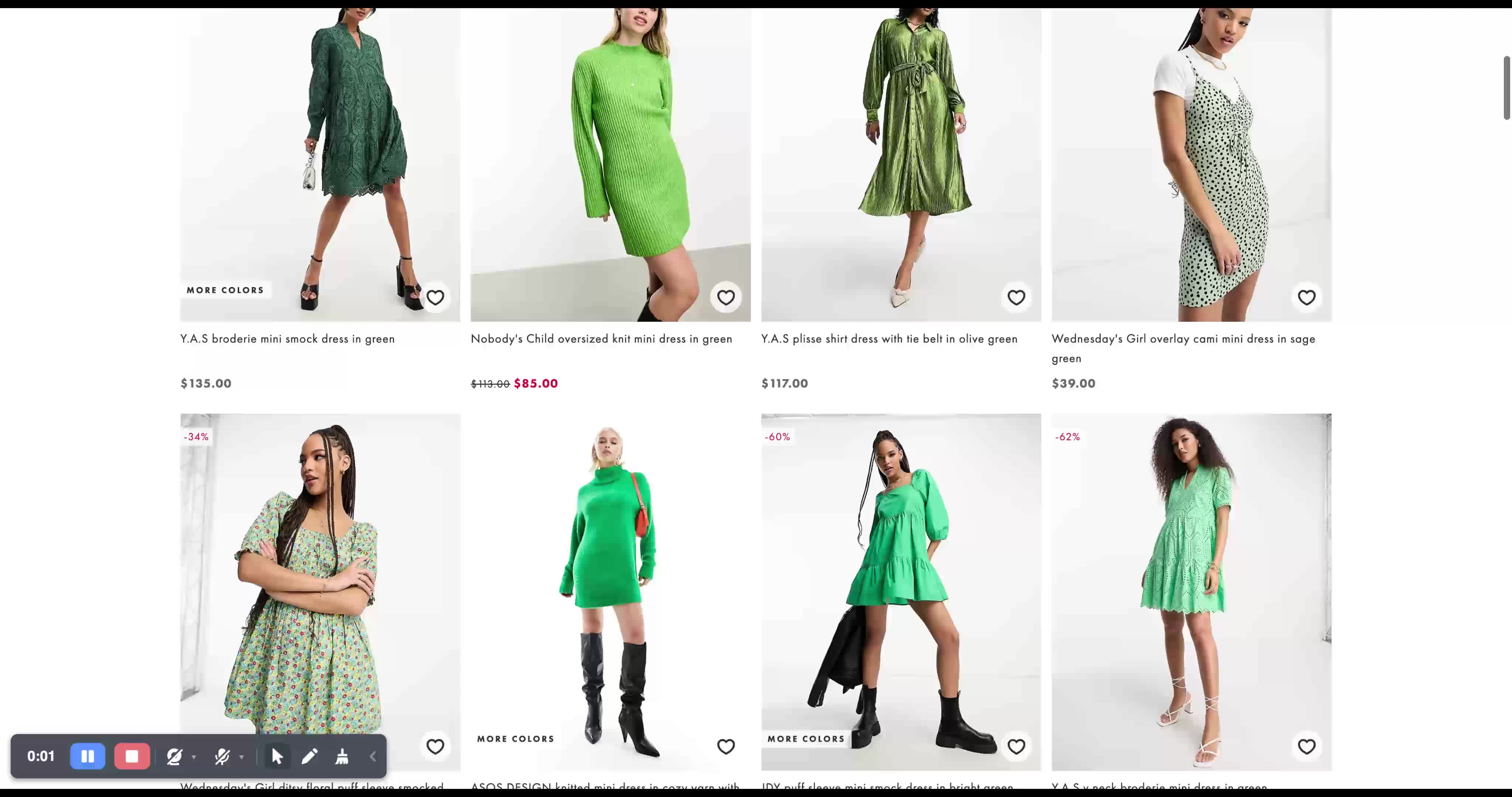}\par
    \includegraphics[width=\linewidth]{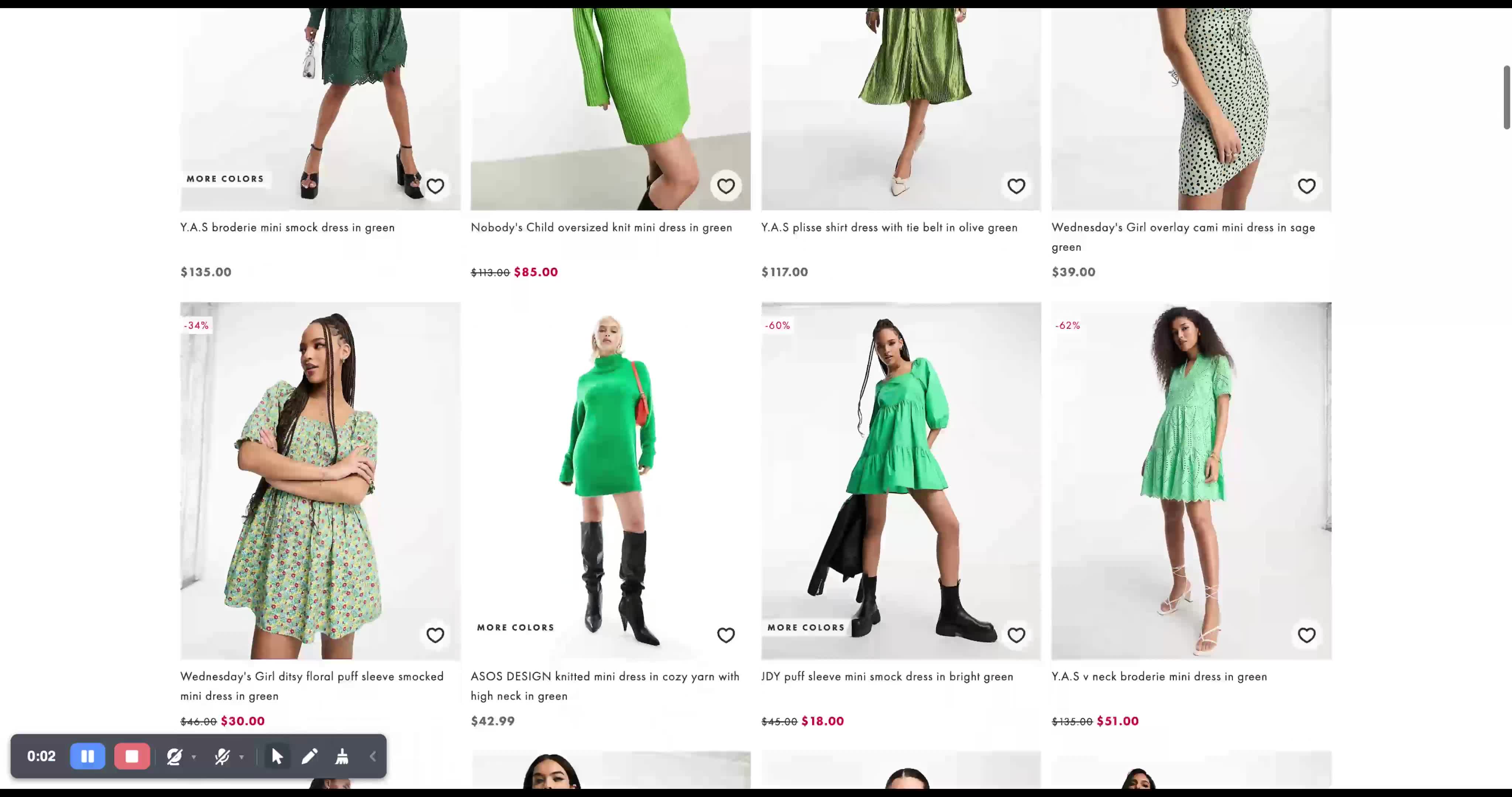}\par
    \includegraphics[width=\linewidth]{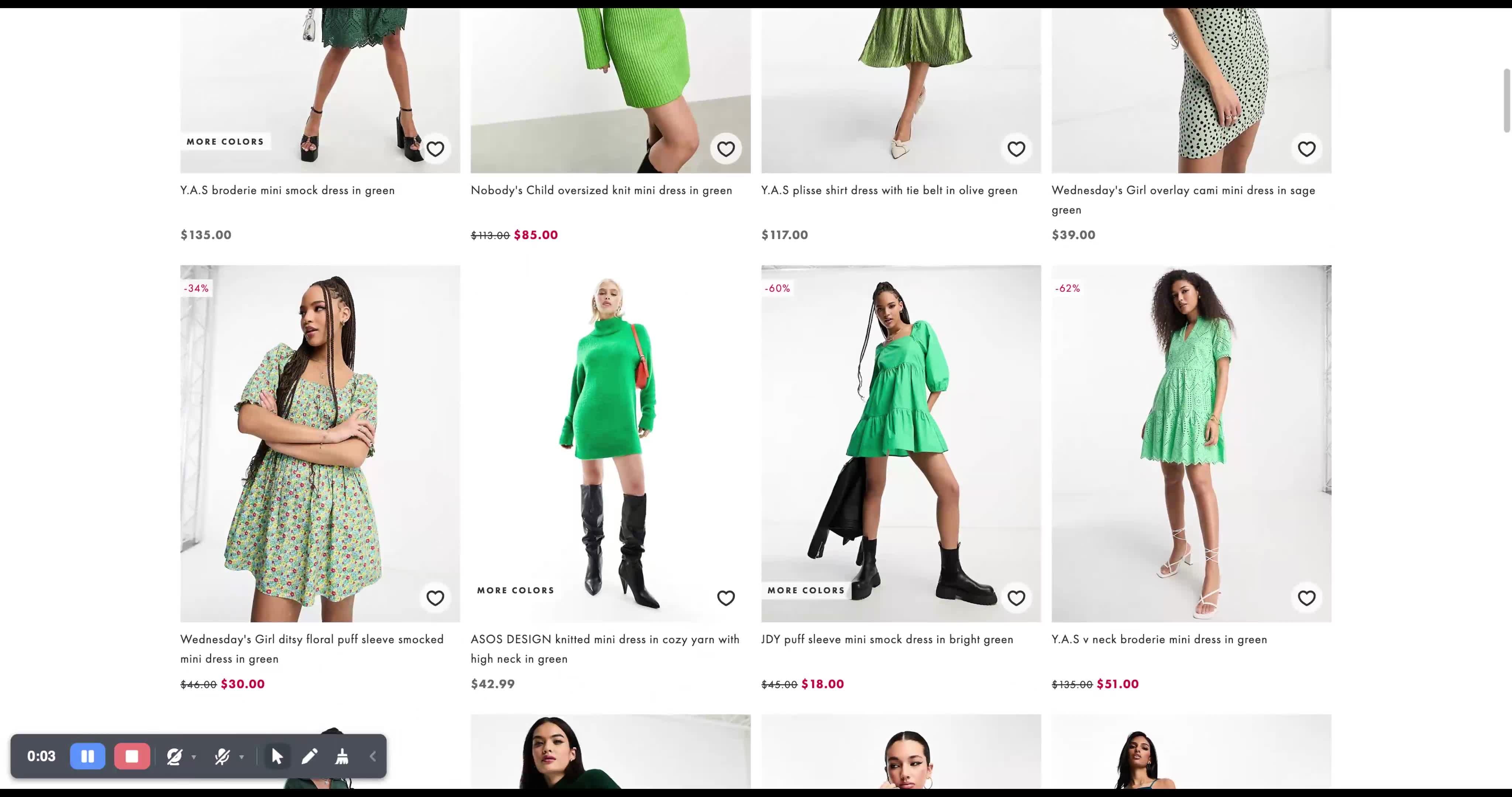}\par
    \includegraphics[width=\linewidth]{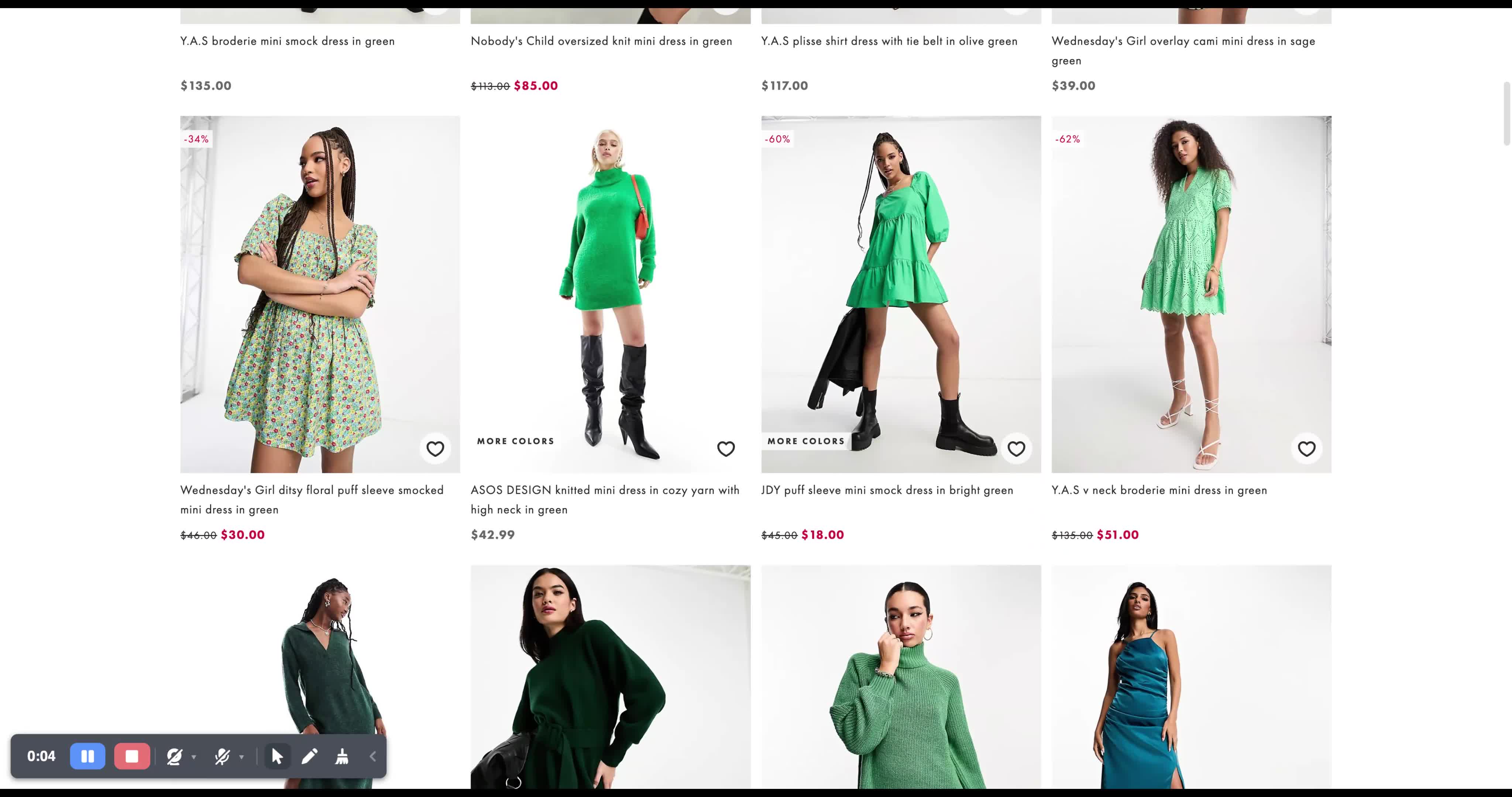}\par
    \includegraphics[width=\linewidth]{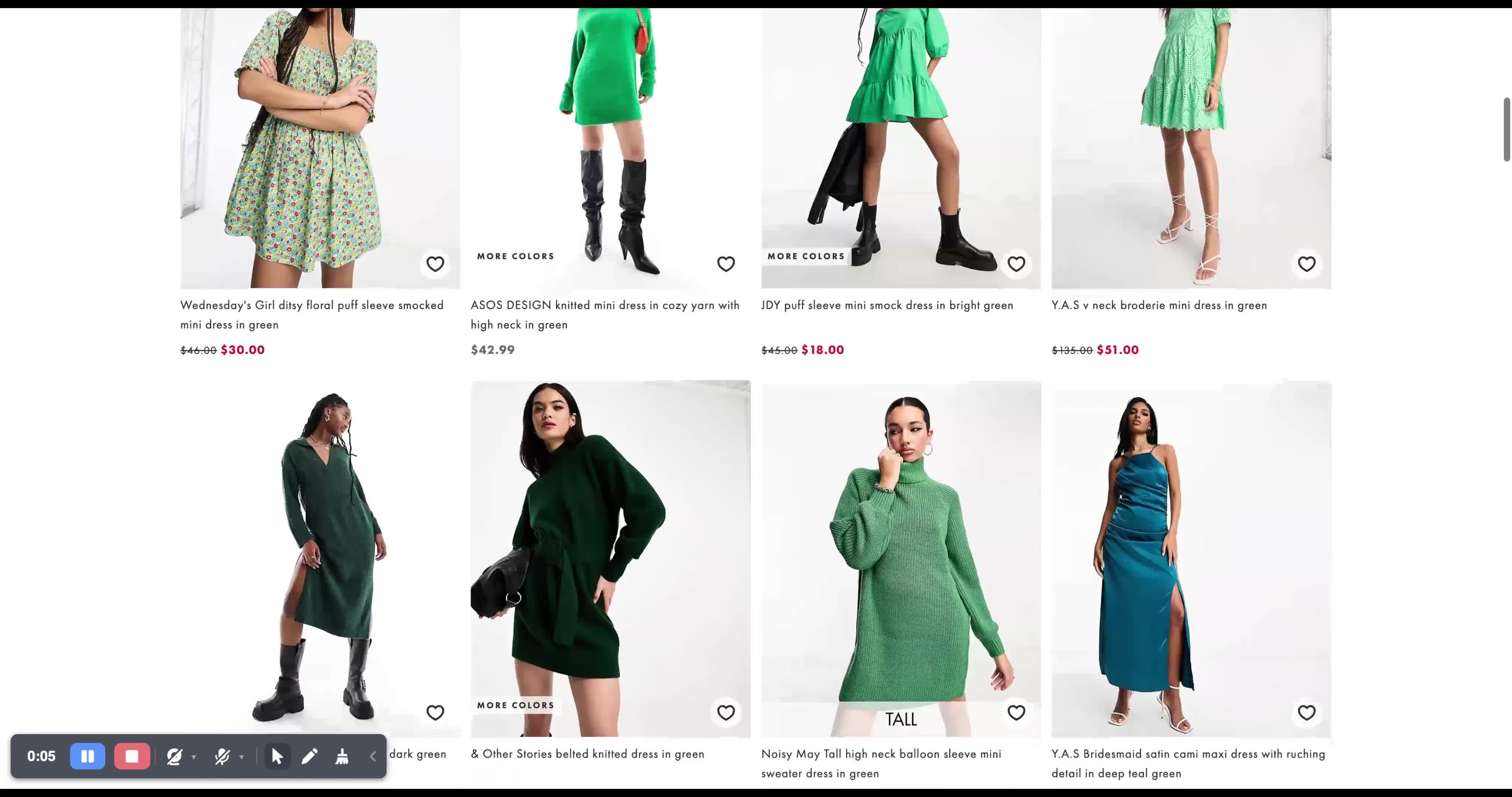}\par
  \end{multicols}
  \caption{Six screen captures showcasing user interactions with the Asos.com website, each captured at a 1-second interval.}
  \label{fig:screen_capture_asos}
\end{figure}

\begin{figure}[htbp]
\begin{tcolorbox}[colframe=cyan, colback=white!10]
 Given the images shown, it's important to note that while certain inferences can be made about product preferences and price sensitivity among other things, these are assumptions based on a set of images featuring green dresses of different styles and prices. Here is a possible interpretation in JSON format: 
``json
\{
  \colorbox{yellow!50}{\textit{``Product Characteristics"}}: ``User appears to be interested in dresses, specifically in shades of \colorbox{yellow!50}{green}, with various styles including mini, midi, and maxi lengths.",
  \colorbox{yellow!50}{\textit{``Lowest Price"}: ``$18.00\$$} for the JDY puff sleeve mini smock dress in bright green",
  \colorbox{yellow!50}{\textit{``Highest Price"}: ``$144.00\$$} for the Y.A.S Bridesmaid satin cami maxi dress with ruching detail in deep teal green",
  \colorbox{yellow!50}{\textit{``Brand Preference"}}: ``Variety of brands shown, such as \colorbox{yellow!50}{Y.A.S, Wednesday's Girl, Nobody's Child,} suggesting no clear preference for a single brand.",
  \textit{``Product Specifications"}: ``Not Available",
  \textit{``User Reviews and Testimonials"}: ``Not Available",
  \textit{``Comparisons"}: ``The user may be comparing dresses based on style and discounts, with several items showing reduced prices.",
  \textit{``Promotions"}: ``Several dresses are listed with percentage discounts, indicating that promotions may influence user preferences."
\}
\end{tcolorbox}
\caption{Extracted summary from $Asos$ website activities revealing a user preference for green dresses, diverse brand interest, and price sensitivity.}
\label{tab:summary_tab_asos}
\end{figure}

\begin{figure}[htbp]
  \centering
  \includegraphics[width=1\linewidth]{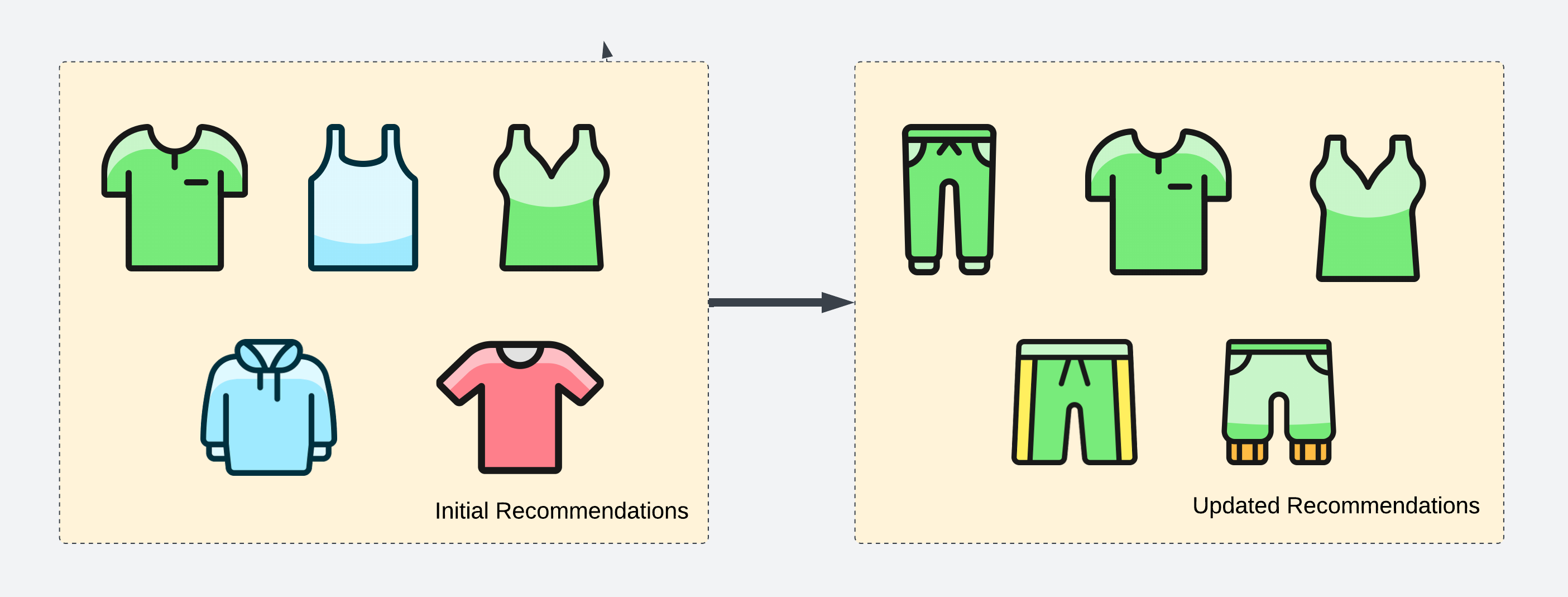}
  \caption{\rec{} Real-time recommendations incorporating the dynamic user behavior.}
  \label{fig:reco}
\end{figure}
We present a dynamic process of generating personalized, real-time recommendations for users navigating the $Asos$ fashion apparel website through an illustrative example. The recommender system (\rec{}) captures screenshots every $2$ seconds, as shown in Figure~\ref{fig:screen_capture_asos}, and stores them in the database. It then processes these screenshots in batches of $10$, due to input token size limitations, and generates keyword summaries based on provided instructions, as illustrated in Figure~\ref{tab:summary_tab_asos} (see Appendix~\ref{app:examples} for more examples). Our approach focuses on extracting constraints like the color and price range of products. \rec{} parses the input, extracting details such as color (`green'), lowest price (`$18\$$'), and highest price (`$144\$$') by leveraging the function-calling ability of the \texttt{gpt-3.5-turbo} model. Using the parsed information as arguments, \rec{} executes the MNL optimization script and communicates the outcomes via the interface. We can observe the impact of incorporating user behavior in the form of updated recommendations being limited to green apparel, as shown in Figure~\ref{fig:reco}. This methodical step-by-step approach ensures consistent refinement of recommendations through continuous analysis of successive batches of images.

\subsection{Experimental validation of \rec{} with session based recommendation model}
\subsubsection{Dataset} 
To validate the \rec{} framework with session-based recommendation model, we need a dataset comprising screenshots of users' browsing sessions. Notably, there is no publicly available dataset meeting these requirements in the existing literature. Therefore, we create a new dataset comprising of screenshots capturing user browsing sessions on the $Amazon.com$ website. Specifically, we leverage the $Amazon-M2$~\citep{jin2024amazon} multilingual shopping session dataset from different regions. The dataset comprises of item IDs from anonymous user browsing sessions in chronological order of their interactions. Using the item IDs we query the $Amazon UK$ website to retrieve the screenshots of corresponding item web pages.

\begin{figure}[htbp]
  \centering
  \includegraphics[width=0.5\linewidth]{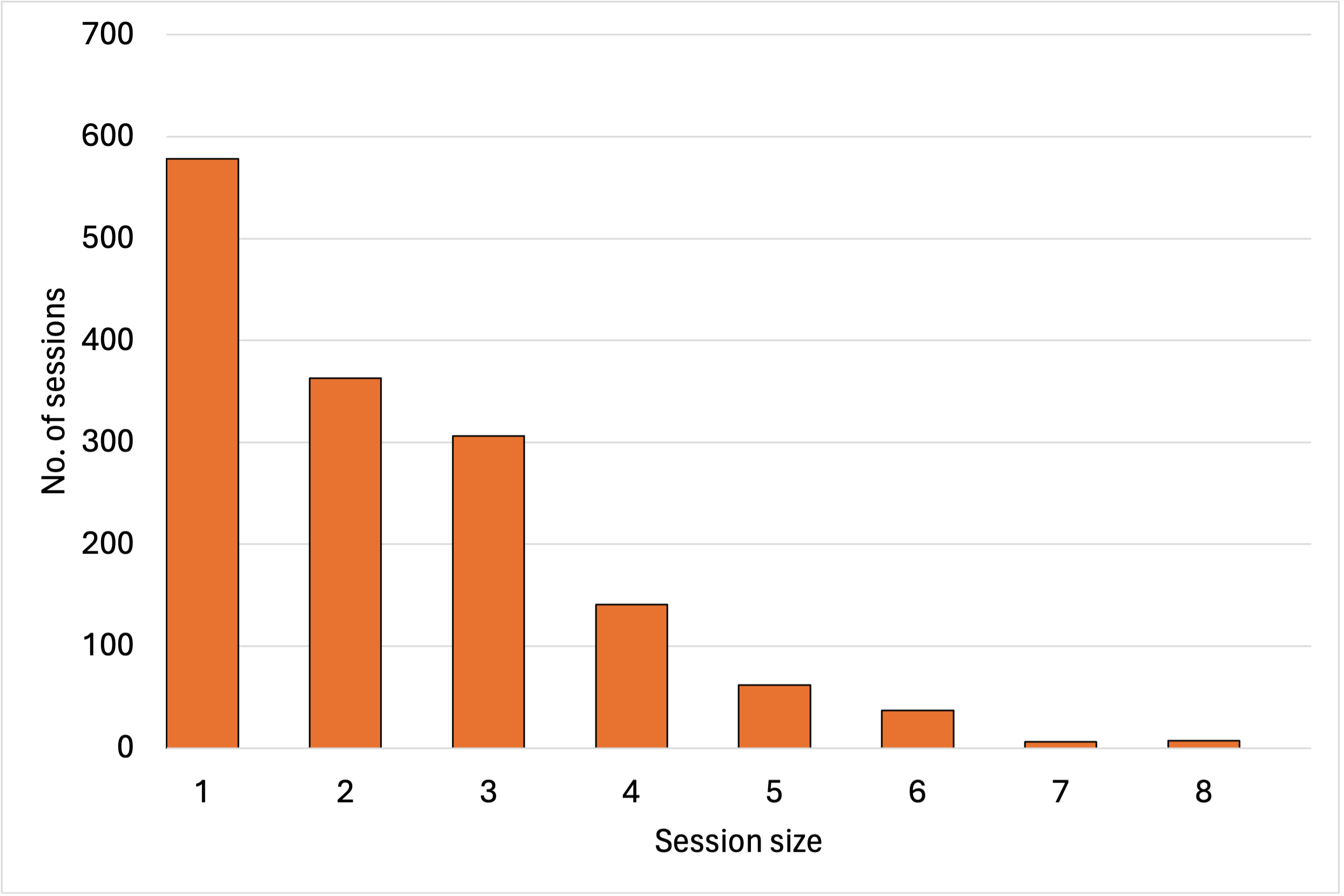}
  \caption{Distribution of session sizes in our session-based screenshot dataset.}
  \label{fig:session-dist}
\end{figure}

\begin{figure}[htbp]
  \begin{multicols}{2}
    \includegraphics[width=\linewidth]{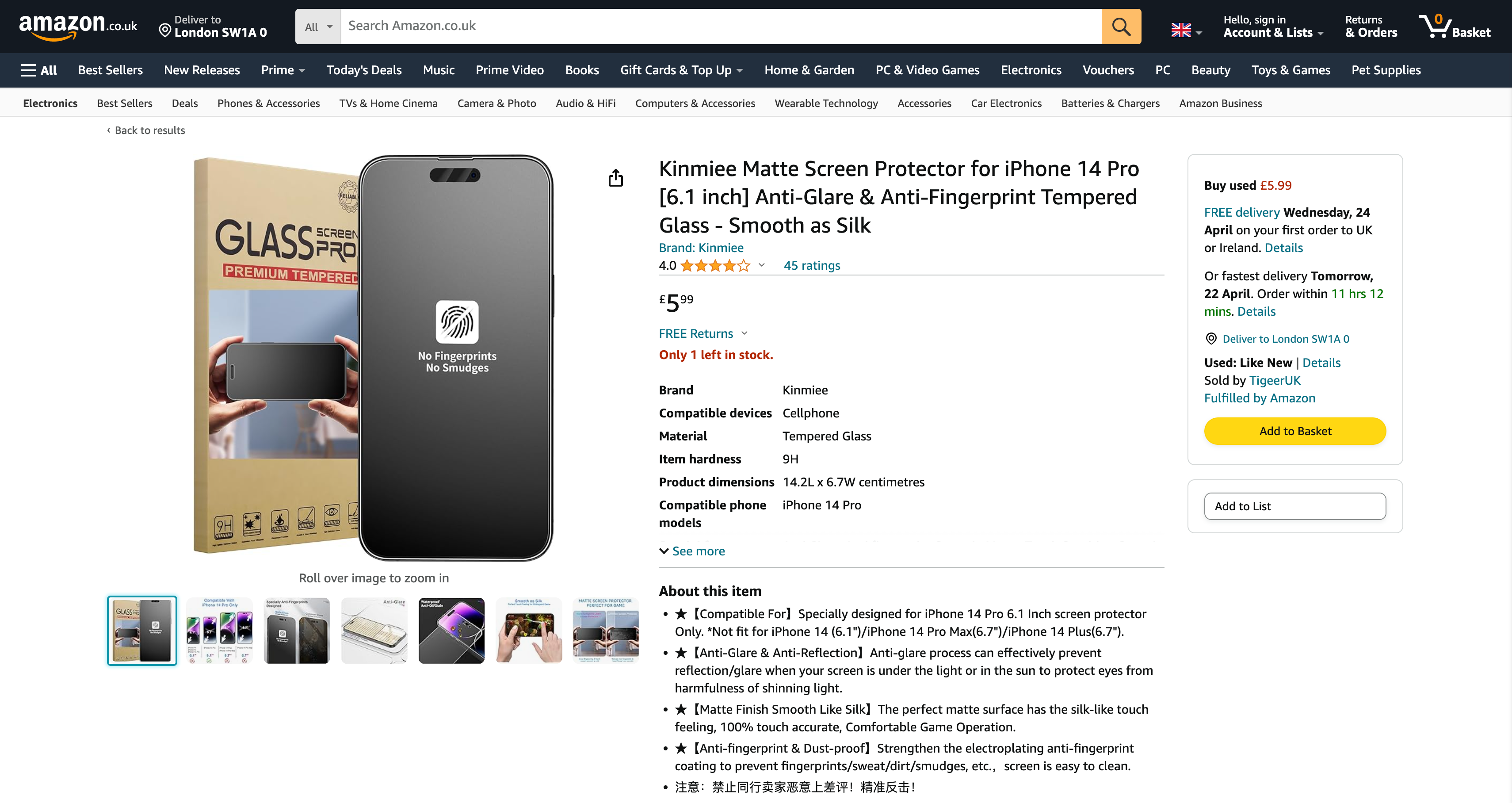}\par
    \includegraphics[width=\linewidth]{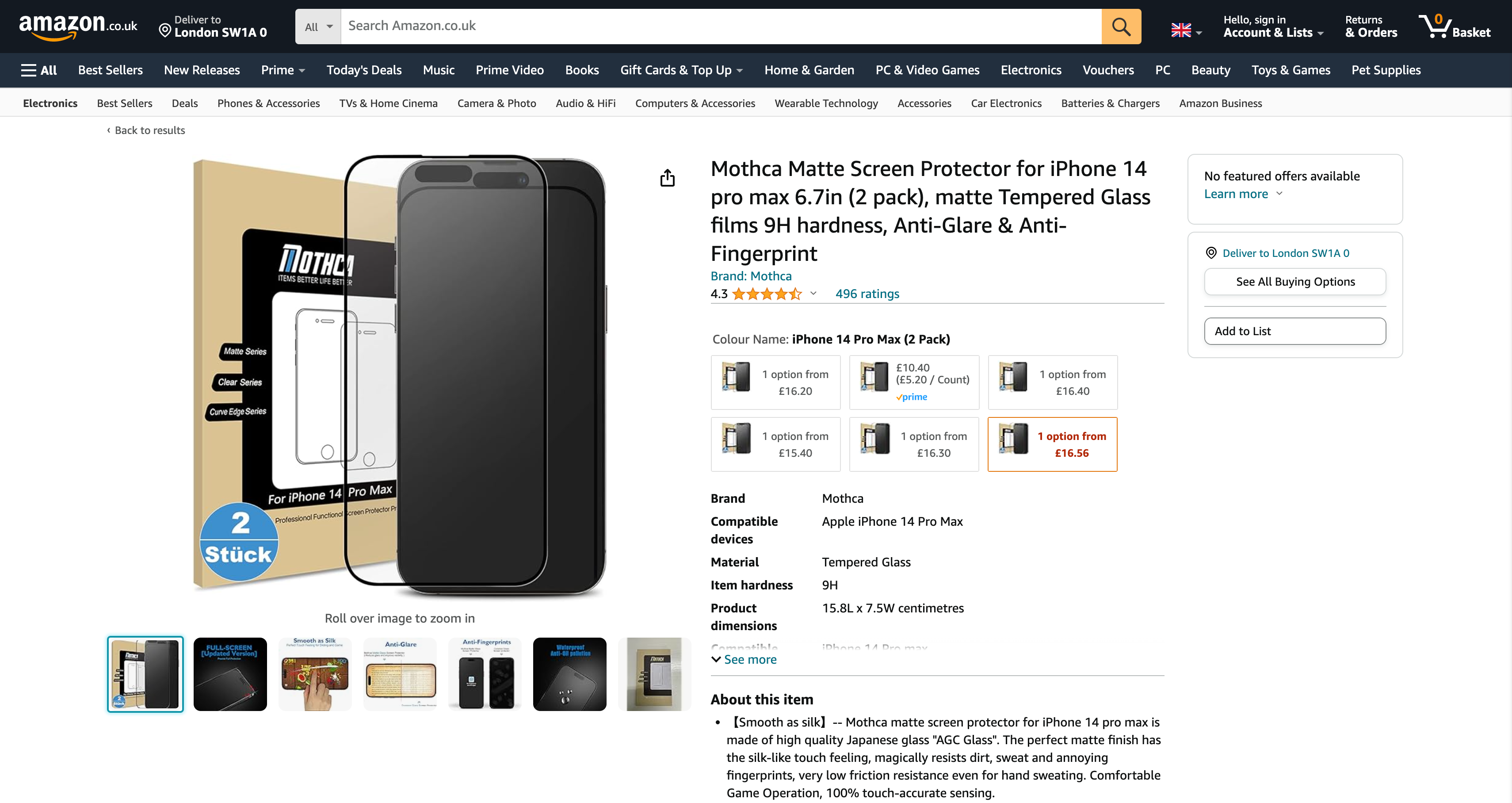}\par
    \includegraphics[width=\linewidth]{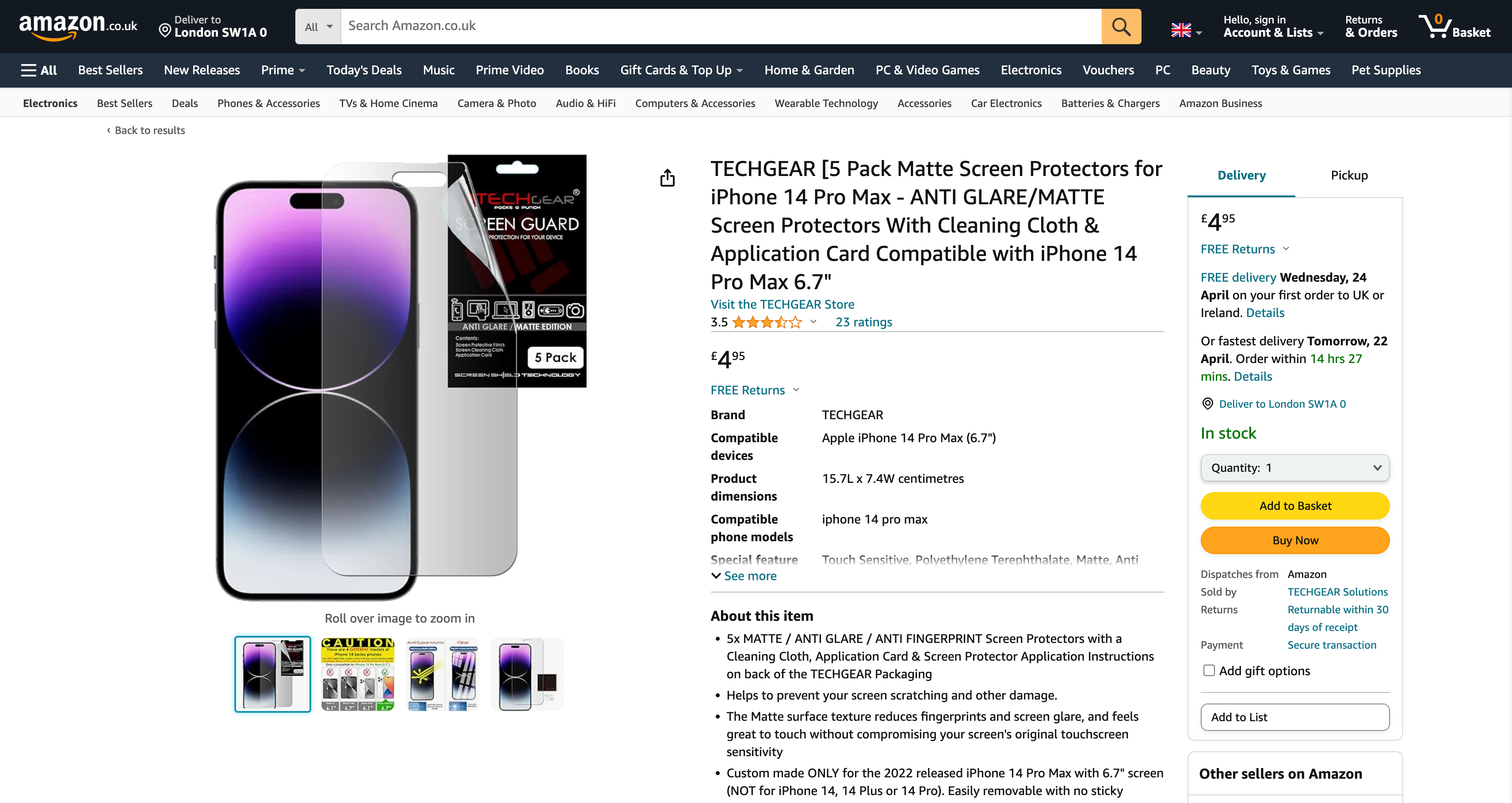}\par
    \includegraphics[width=\linewidth]{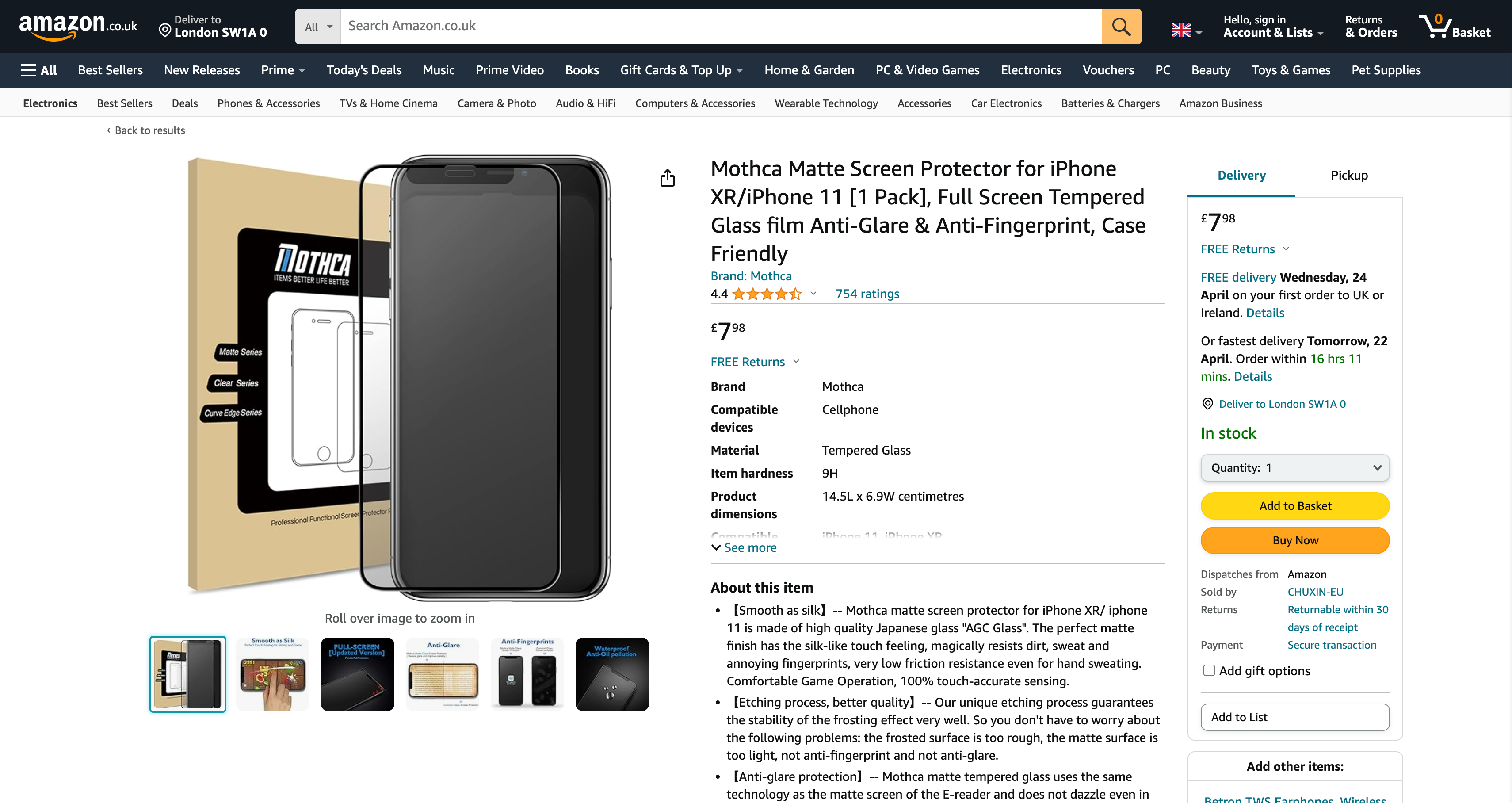}\par
  \end{multicols}
  \caption{Screenshots from a user browsing session on the Amazon website.}
  \label{fig:screens}
\end{figure}

The resulting dataset contains screenshot of item web pages from $1,500$ sessions. Figure~\ref{fig:session-dist} details the distribution of screenshots per session, showing that the majority of sessions contain fewer than five items. Figure~\ref{fig:screens} illustrates the screenshots of a user browsing for screen protectors in a session. These screenshots accurately capture all the information visible to the user during their interactions.

\subsubsection{Popular baselines}
We evaluate the performance of several popular sequential session-based recommendation models, both in their original form and when integrated into the proposed \rec{} framework. 
\begin{itemize}
    \item \textbf{HRM}~\citep{wang2015learning} uses a hierarchical representation model to model user sessions.
    \item \textbf{GRU4Rec}~\citep{tan2016improved} uses a recurrent neural network to model user sessions.
    \item \textbf{NARM}~\citep{li2017neural} employs a hybrid encoder with an attention mechanism to model user sessions.
    \item \textbf{NPE}~\citep{nguyen2018npe} uses a neural personalized embedding model to model user sessions.
    \item \textbf{SR-GNN}~\citep{wu2019session} uses graph neural networks to model user sessions.
\end{itemize}

\subsubsection{Evaluation metrics}
We evaluate the quality of the recommendations using two widely-adopted rank-based metrics: Recall@k and MRR@k. Recall@k measures the proportion of instances where the relevant item is among the top-k recommended items, while MRR@k computes the average of the inverse ranks of the relevant items.

\subsubsection{Implementation details}
\paragraph{Training setup:}
We train the baseline models using the Pytorch on a Linux machine with a $12$ core, $32$GB 64-bit Intel(R) Core(TM) i7-8700 CPU @ 3.20GHz processor, and an Nvidia GeForce GTX 1080 GPU. We use 64-dimensional embeddings to represent items. The models are optimized using Adam~\cite{kingma2014adam} with an initial learning rate of 0.001 and a mini-batch size of 512. Hyper-parameters are tuned on the validation set, and training is terminated if the validation performance doesn't improve for 10 epochs. Finally, we report the performance of each method under its optimal hyper-parameter settings. 

\paragraph{Reranking setup:} 
We utilize the \texttt{SentenceTransformers} framework to create sentence embeddings and employ \texttt{gpt-4-vision-preview} (GPT-4V) as our MLLM to process screenshots and generate keyword summaries.

\subsubsection{Results}
\begin{figure}[htbp]
  \begin{multicols}{2}
    \includegraphics[width=\linewidth]{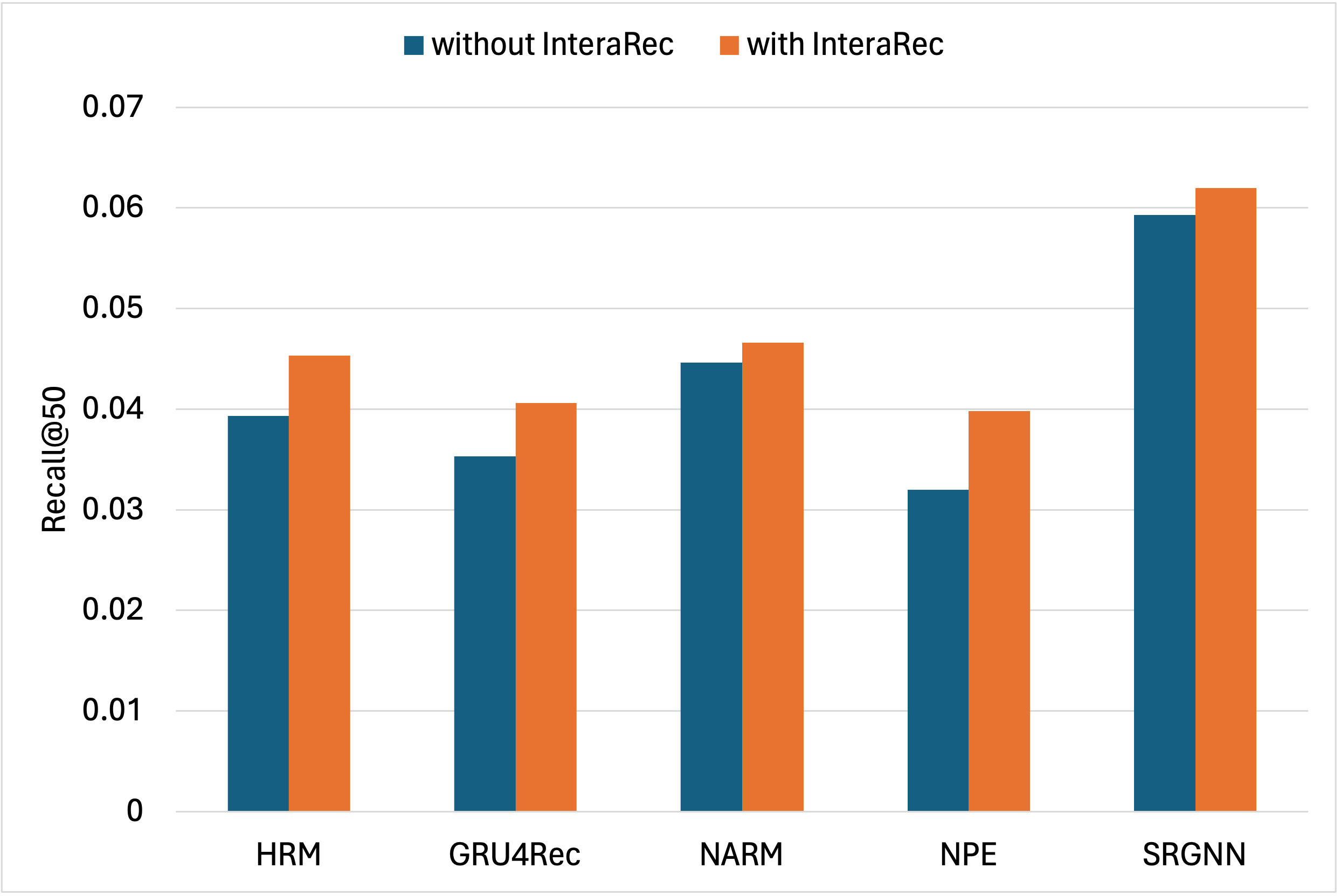}\par
    \includegraphics[width=\linewidth]{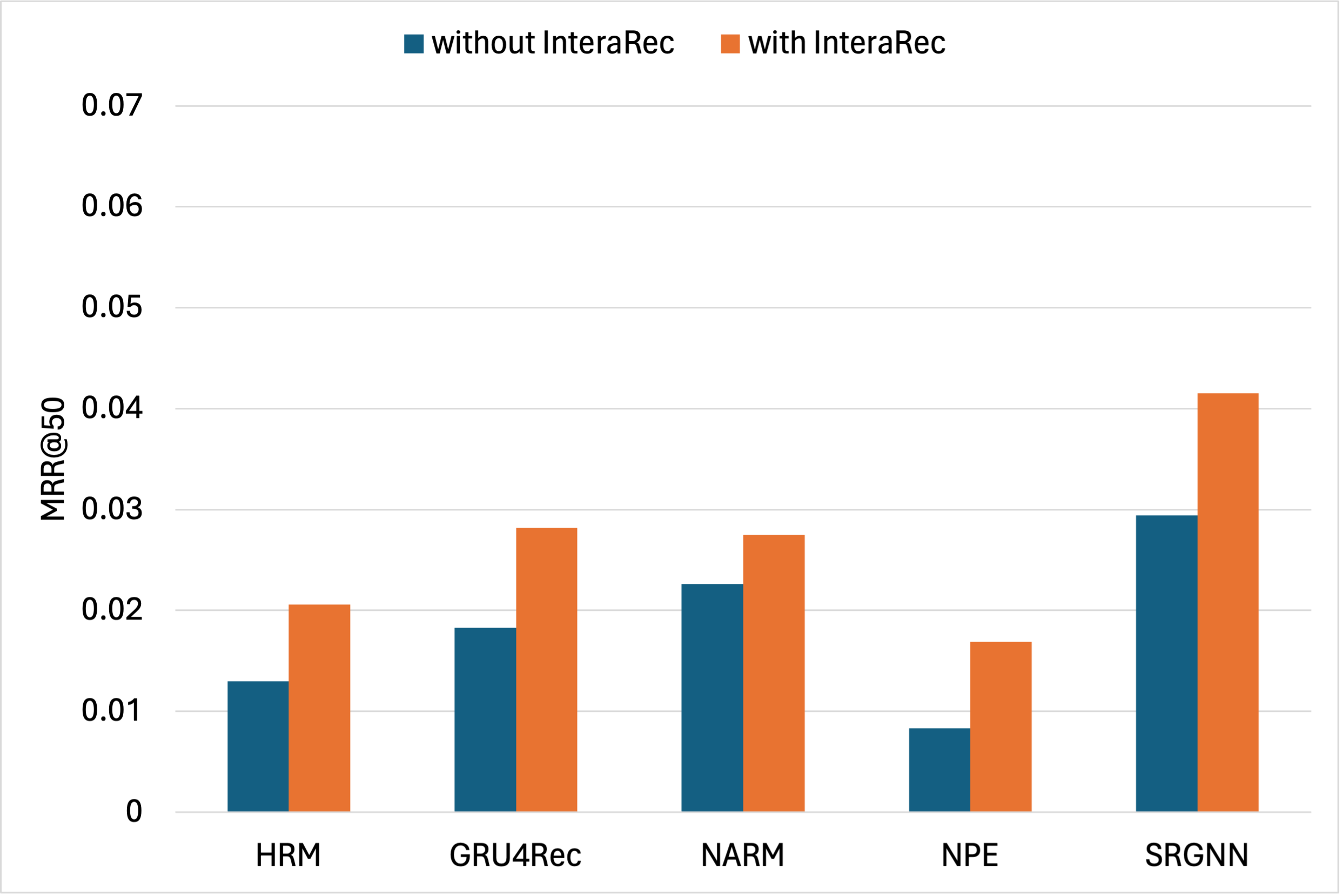}\par
  \end{multicols}
  \caption{Ranking performance of baseline models with and without \rec{}.}
  \label{fig:baseline_improvemnts}
\end{figure}

We observe performance gains across all baselines using the \rec{} framework, as indicated by the evaluation metrics MRR@50 and Recall@50, shown in Figure~\ref{fig:baseline_improvemnts}. This demonstrates the effectiveness of the \rec{} framework in enhancing prediction rankings by incorporating additional context from keyword summaries of user browsing session screenshots.

\begin{figure}[htbp]
  \begin{multicols}{2}
    \includegraphics[width=\linewidth]{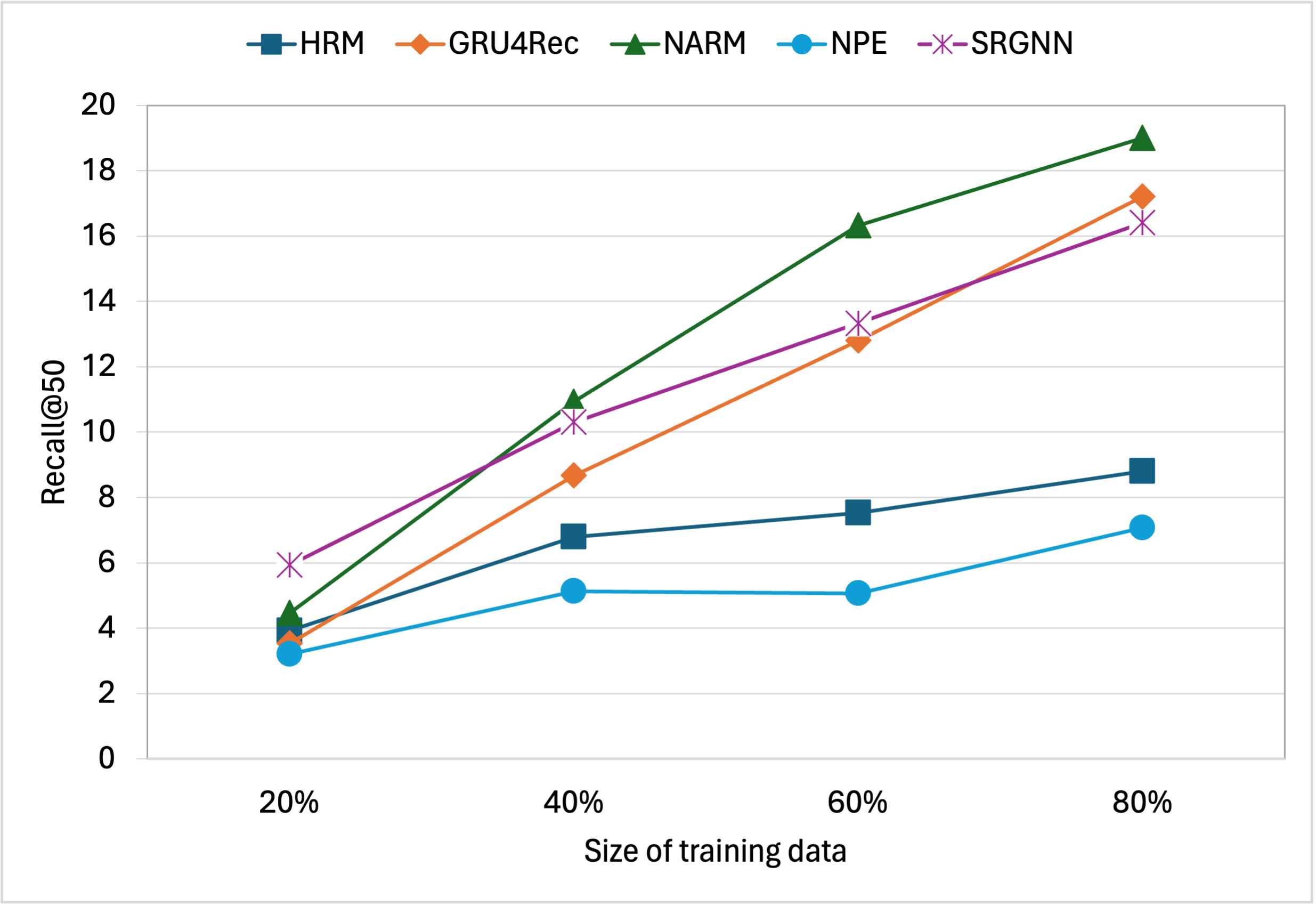}\par
    \includegraphics[width=\linewidth]{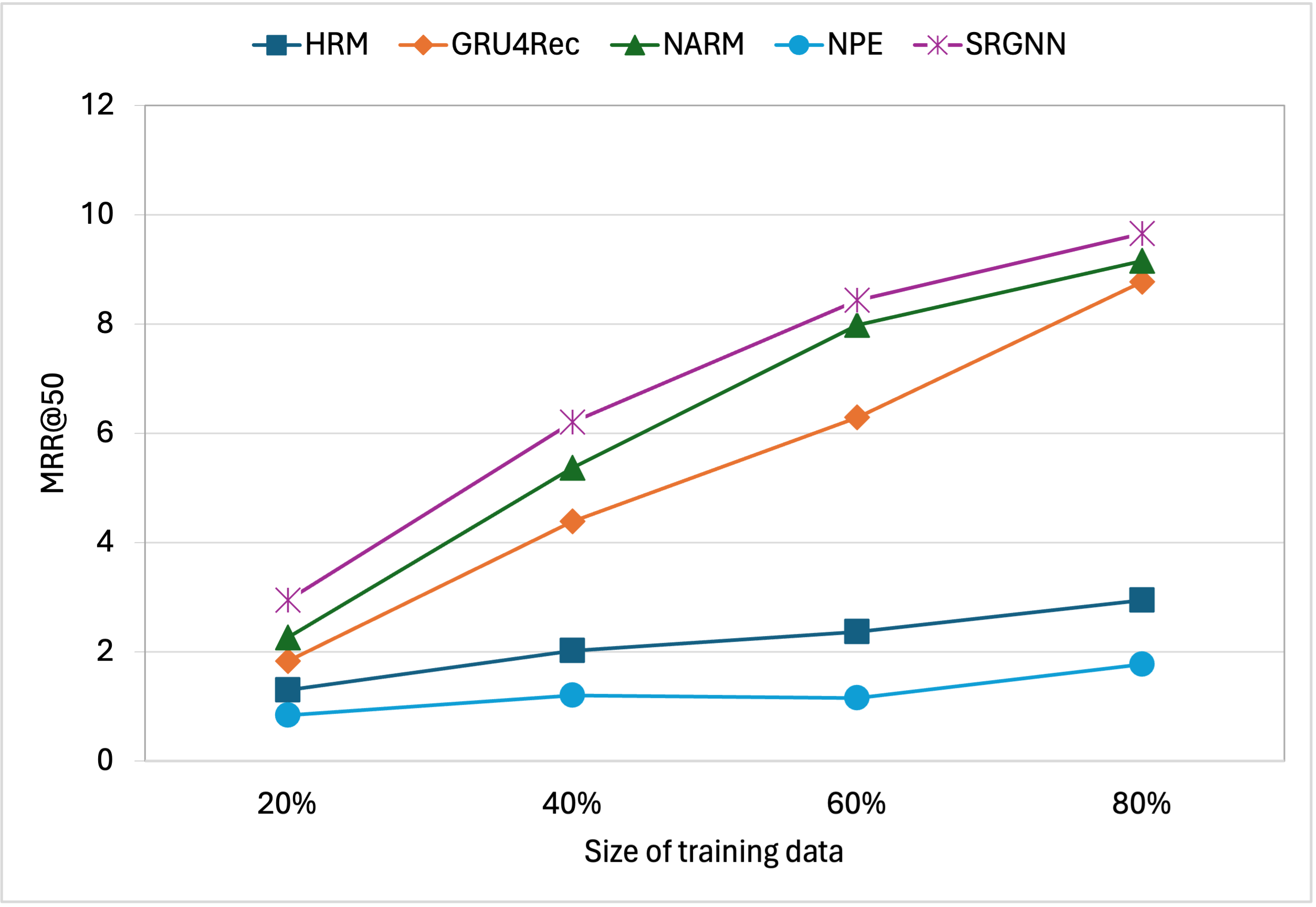}\par
  \end{multicols}
  \caption{Impact of augmenting the training data on ranking performance of baseline models.}
  \label{fig:train_baseline_improvemnts}
\end{figure}

\begin{figure}[htbp]
\begin{multicols}{2}
    \begin{center}
        \includegraphics[width=\linewidth]{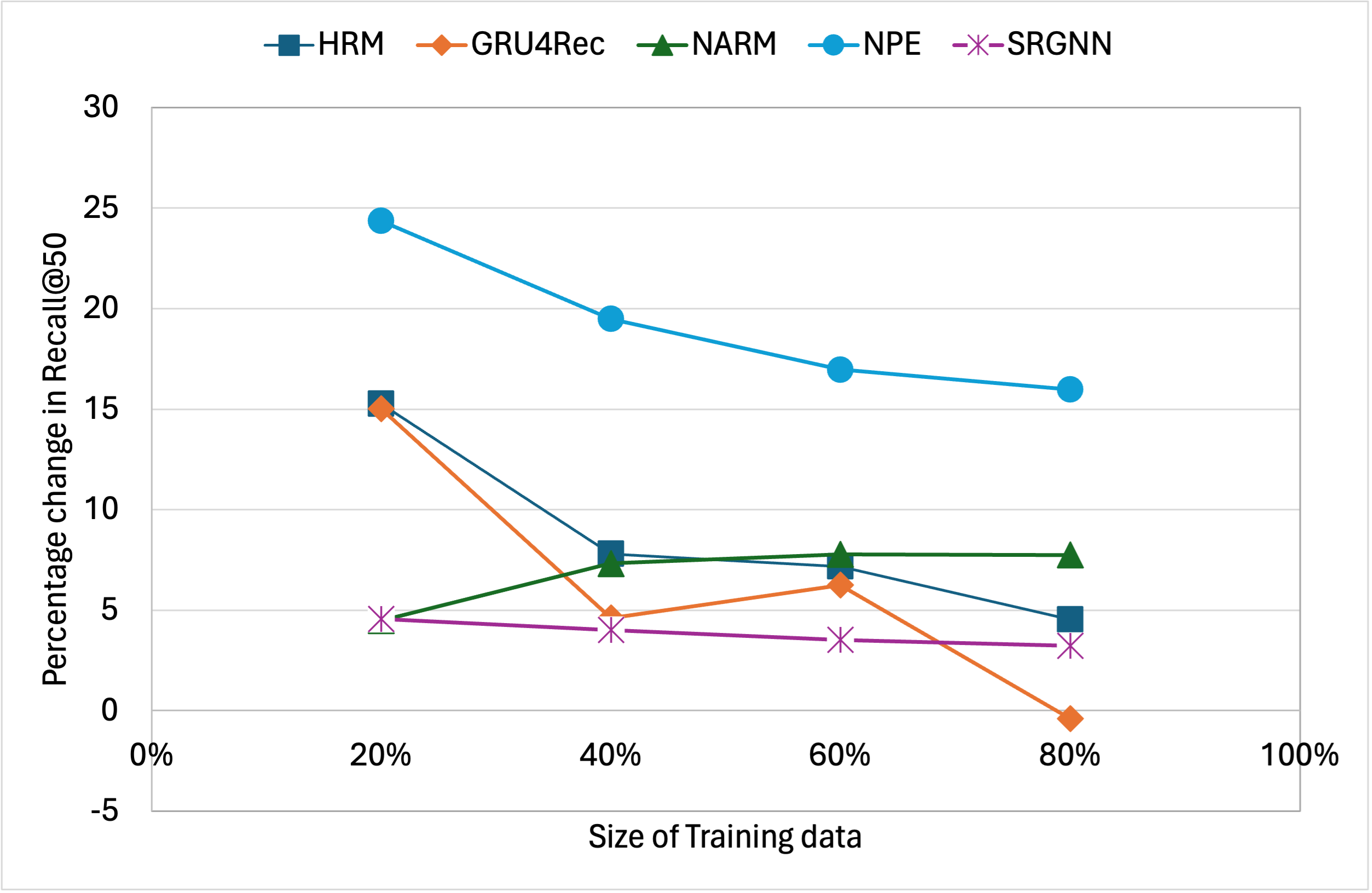}
    \end{center}
    \columnbreak
    \begin{center}
        \includegraphics[width=\linewidth]{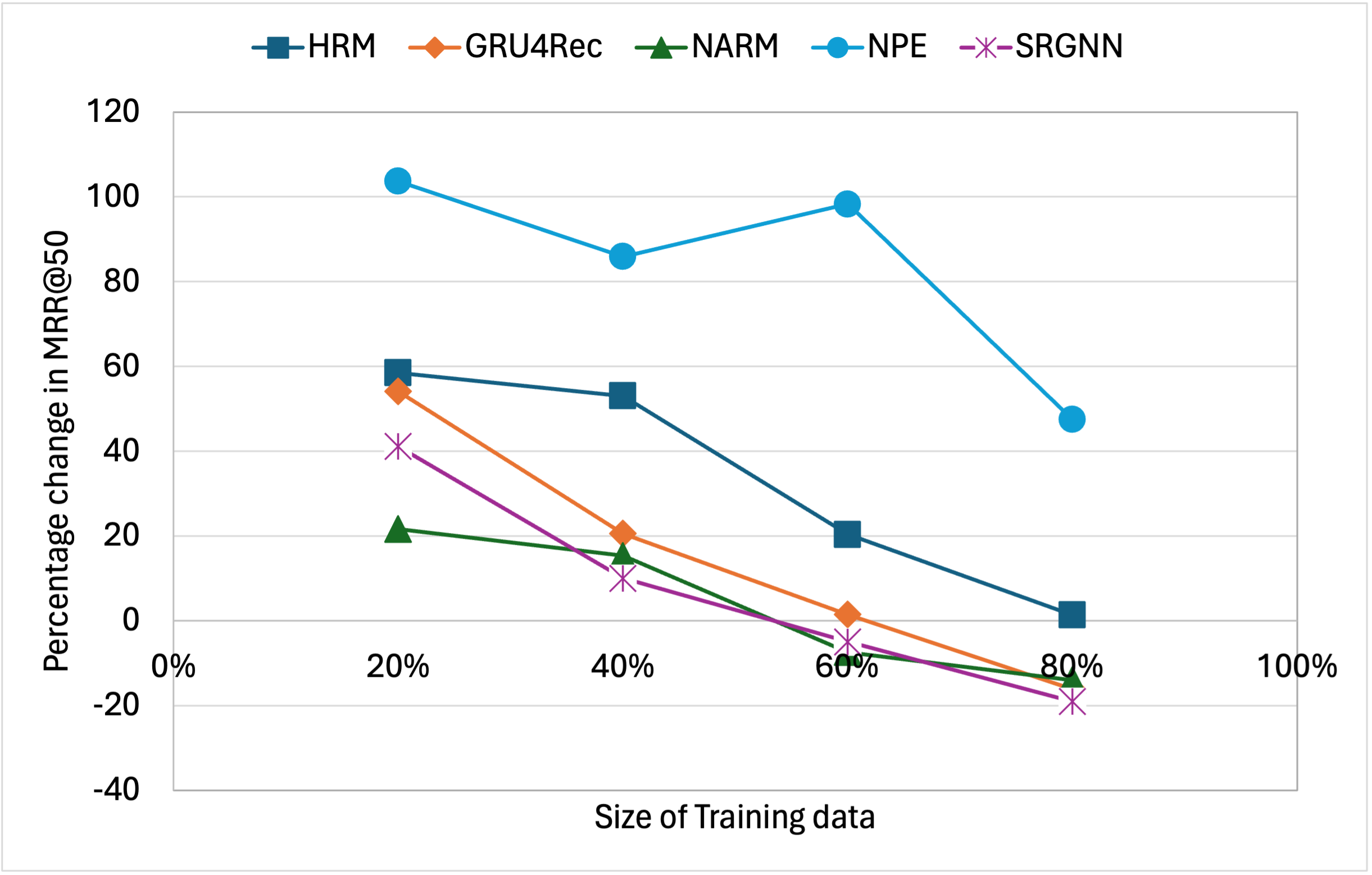}
    \end{center}
\end{multicols}
    \caption{Impact of augmenting the training data on ranking performance of baseline models with \rec{} framework.}
    \label{fig:compare_baseline_improvements}
\end{figure}    

\paragraph{Effects of augmenting training data}
Expanding on the previous findings, we conducted a series of experiments where we incrementally augmented the size of the training dataset for all baseline models. Our results, illustrated in Figure~\ref{fig:train_baseline_improvemnts}, demonstrate that the performance of all baseline models improved as the training data size grew. However, we observed a notable reduction in performance across all baselines when employing the \rec{} framework as shown in Figure~\ref{fig:compare_baseline_improvements}. This suggests that the influence of the keyword summary on re-ranking predictions diminishes as the models become more robust. This phenomenon can be attributed to the enhanced baselines' ability to indirectly capture and incorporate the information from the keyword summary into their predictions, owing to the larger training dataset.

\begin{figure}[htbp]
  \begin{multicols}{2}
    \includegraphics[width=\linewidth]{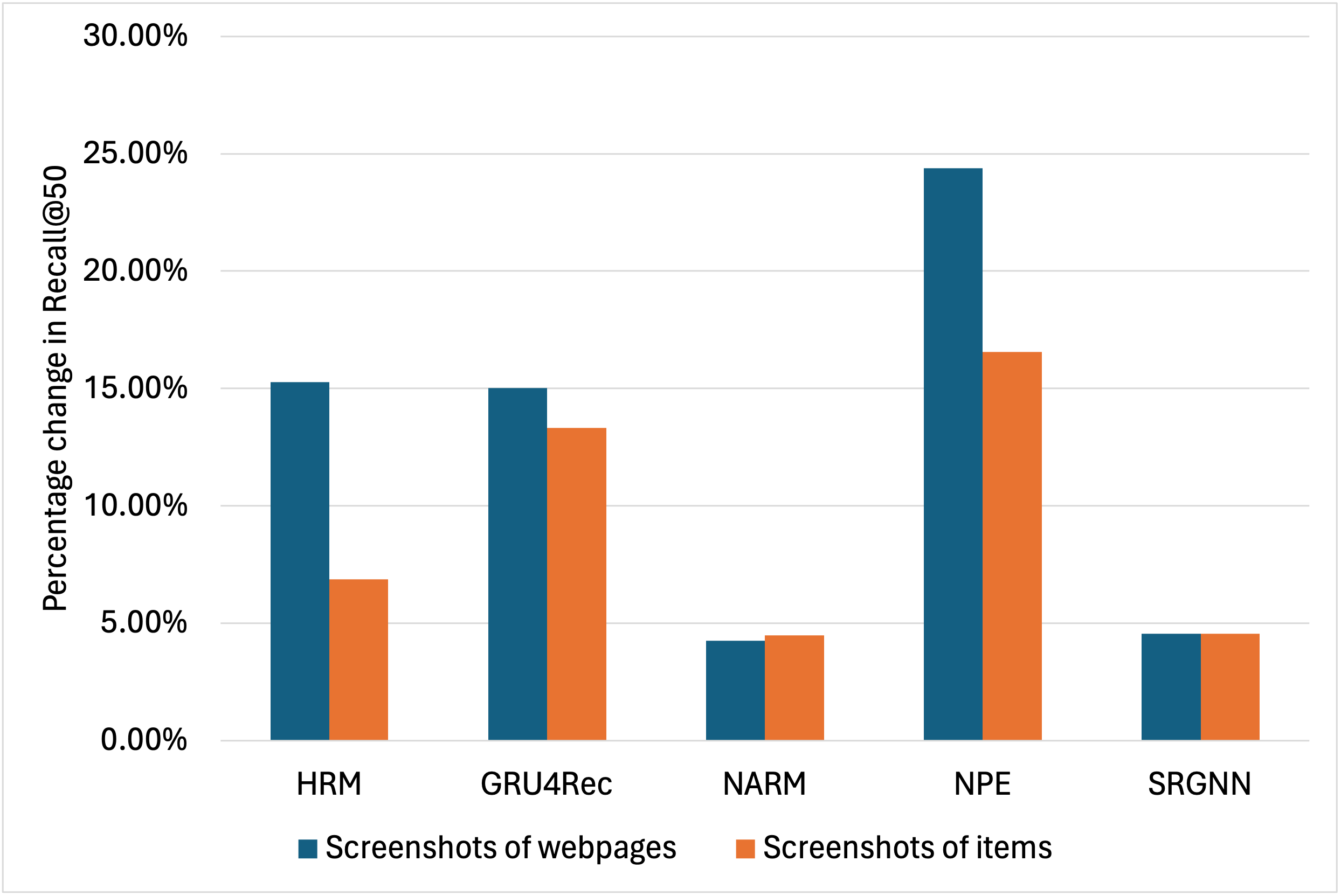}\par
    \includegraphics[width=\linewidth]{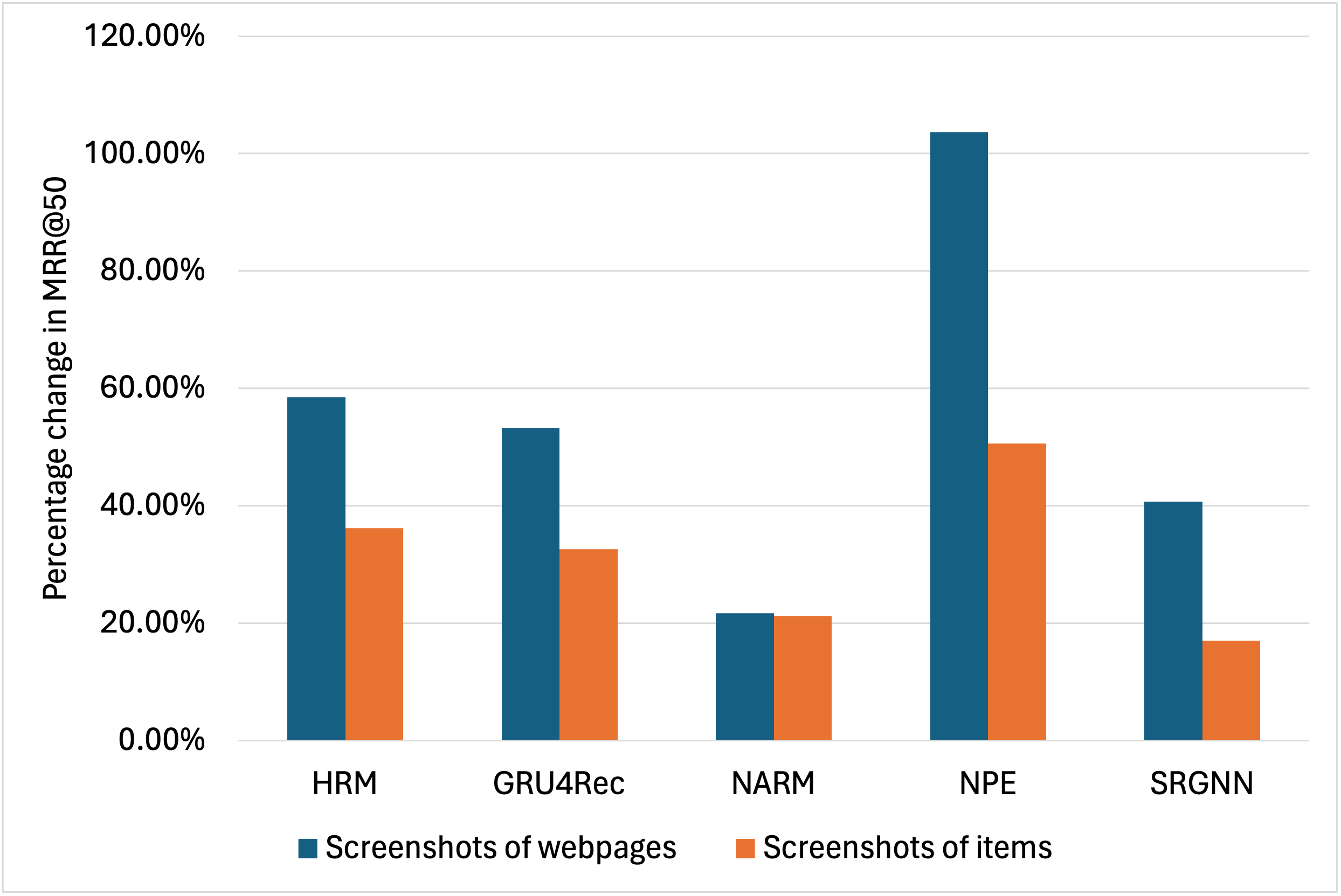}\par
  \end{multicols}
  \caption{Impact of screenshot type on ranking performance of baseline models with \rec{} framework.}
  \label{fig:pvs}
\end{figure}

\paragraph{Effects of exclusively using product images}
In this set of experiments, we leveraged solely the screenshots of individual items from web pages, rather than entire item web pages, as input to the \rec{} framework for generating keyword summaries representing user interests and preferences. The corresponding results are presented in Figure~\ref{fig:pvs}. We observe a decrease in overall performance, as item images provide limited context compared to web pages, which contain additional information such as item titles, brands, and other specifications. This suggests that the lack of this additional data hindered the ability to generate relevant summaries, leading to a performance decline. Additionally, we observe that MLLM was able to filter out any additional noise within the screenshots of web pages. 

\begin{figure}[htbp]
  \begin{multicols}{2}
    \includegraphics[width=\linewidth]{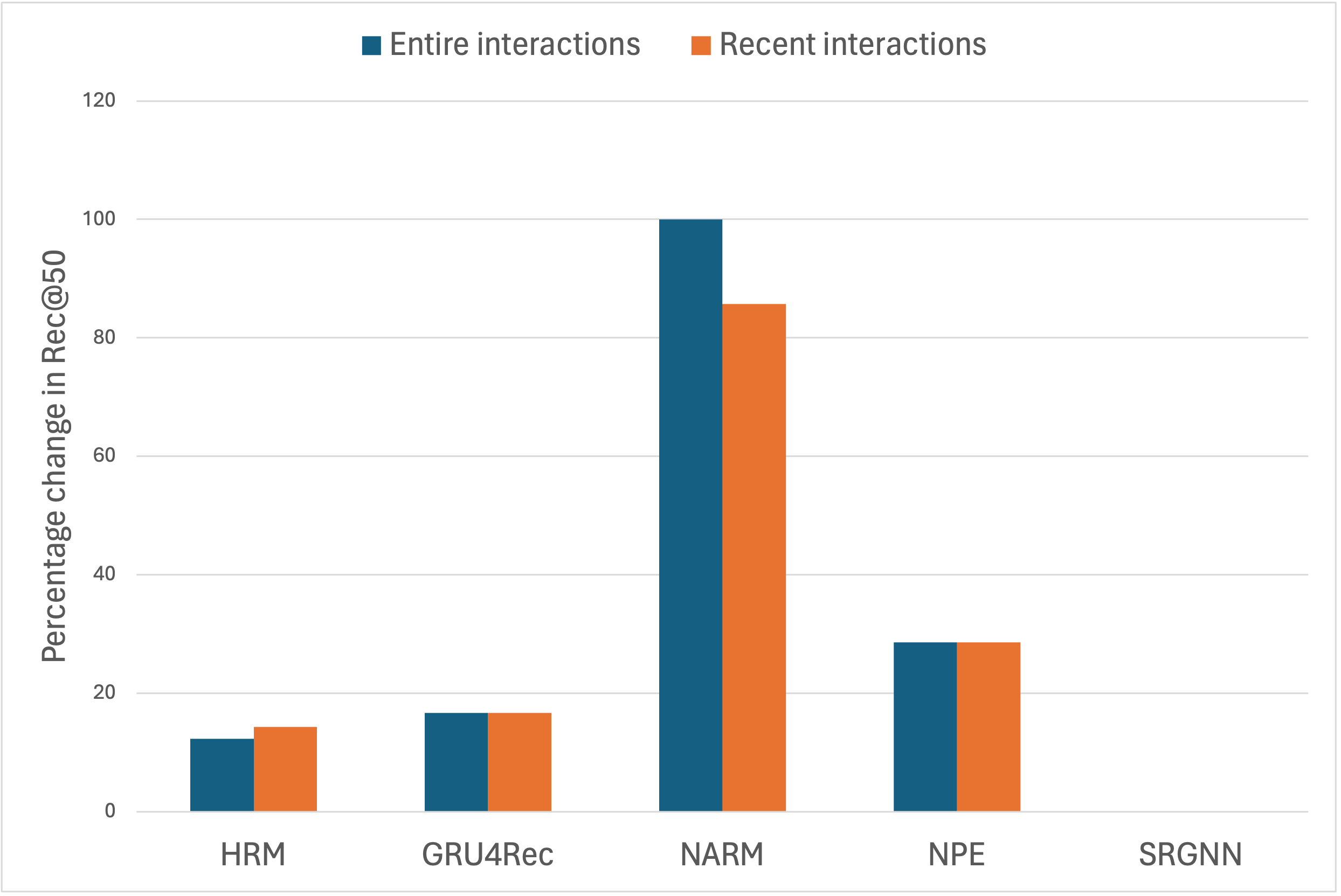}\par
    \includegraphics[width=\linewidth]{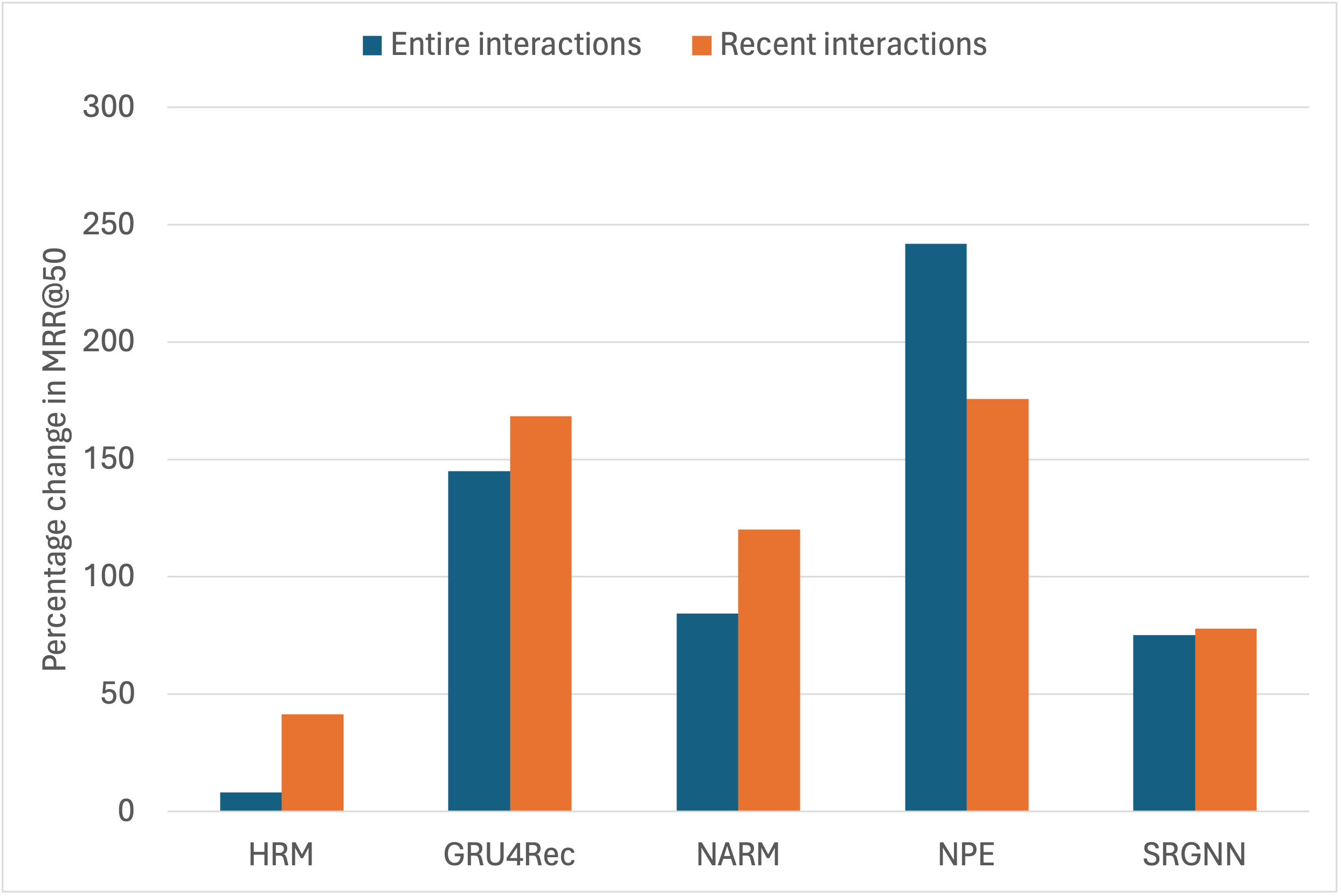}\par
  \end{multicols}
  \caption{Impact of limiting the session size on ranking performance of baseline models with \rec{} framework.}
  \label{fig:ssvs}
\end{figure}

\paragraph{Effects of limiting the session size}
In this set of experiments, we focus on generating summaries by leveraging the recent history of interactions from a given session and compare the performance improvement to the setting where we leveraged the entire history of interactions from a session. Our results showed an improvement in performance across all baselines except for the NPE model as shown in the Figure~\ref{fig:ssvs}. This suggests that recent interactions have a significant influence on overall performance of the baselines, whereas using the entire interaction history may reduce the impact of recent interactions by treating all interactions with equal importance, rather than prioritizing the most recent ones.

\section{Conclusion} \label{sec:conclusion}

In this paper, we introduce \rec{}, an interactive framework designed to craft personalized recommendations for users browsing on an e-commerce platform. Overcoming the limitations of extensive weblog engineering, \rec{} utilizes screenshots to capture and understand user behavior. Harnessing the capabilities of both LLM and MLLMs, \rec{} leverages insights from the screenshots and intelligently executes suitable optimization tools, translating user behavior into concise, easily interpretable recommendations for users. Furthermore, our findings demonstrate that integrating state-of-the-art session-based recommendation systems with the \rec{} framework enhances their performance. Our findings underscore the substantial potential of LLMs in the realm of recommender systems. Overall, \rec{} heralds a new era in recommendation systems driven by visual data.

\bibliography{references}

\newpage
\appendix
\section{} \label{app:examples}
We present some more illustrative example of \rec{} framework below.
\begin{figure}[htbp]
  \begin{multicols}{3}
    \includegraphics[width=\linewidth]{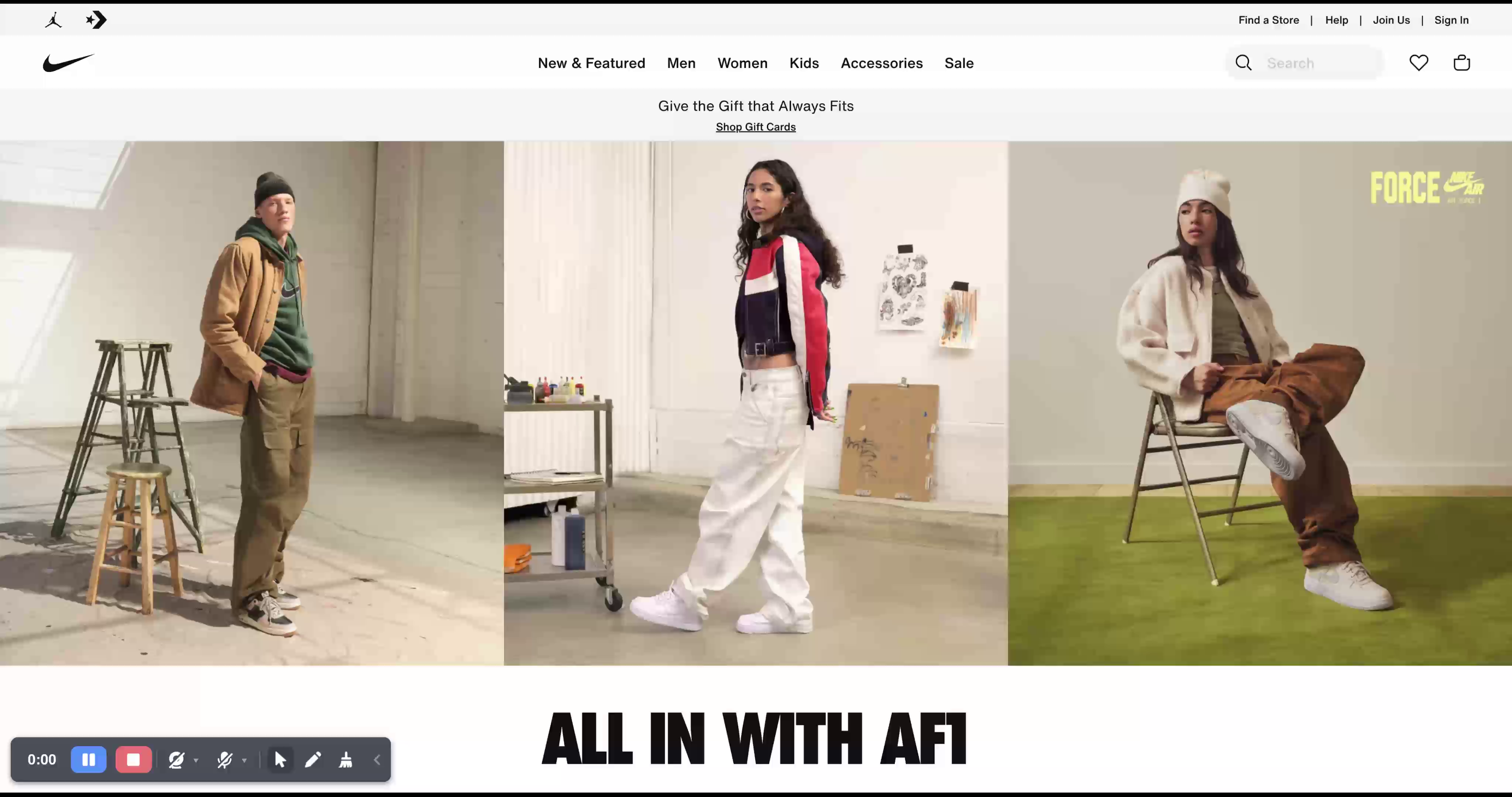}\par
    \includegraphics[width=\linewidth]{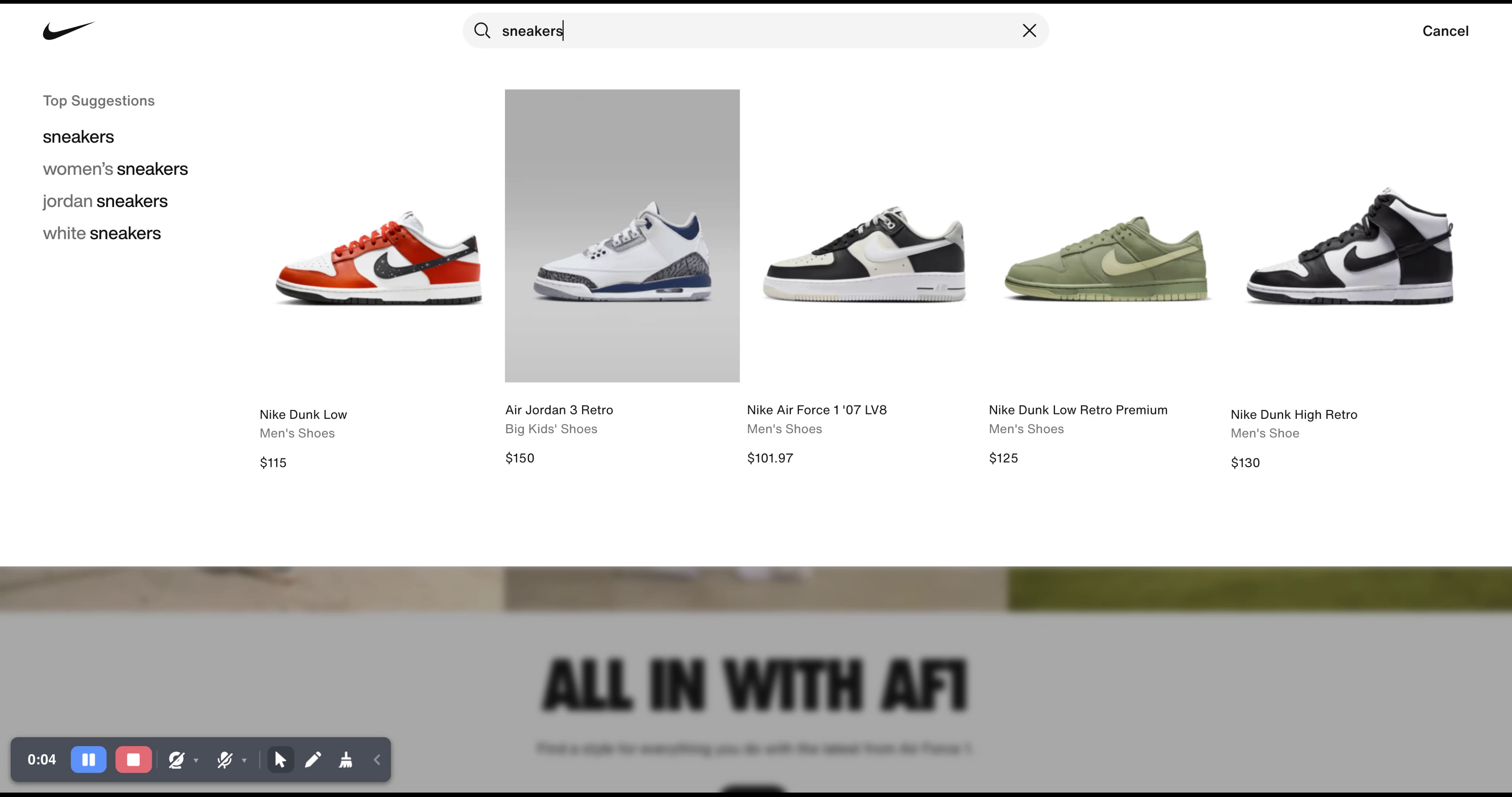}\par
    \includegraphics[width=\linewidth]{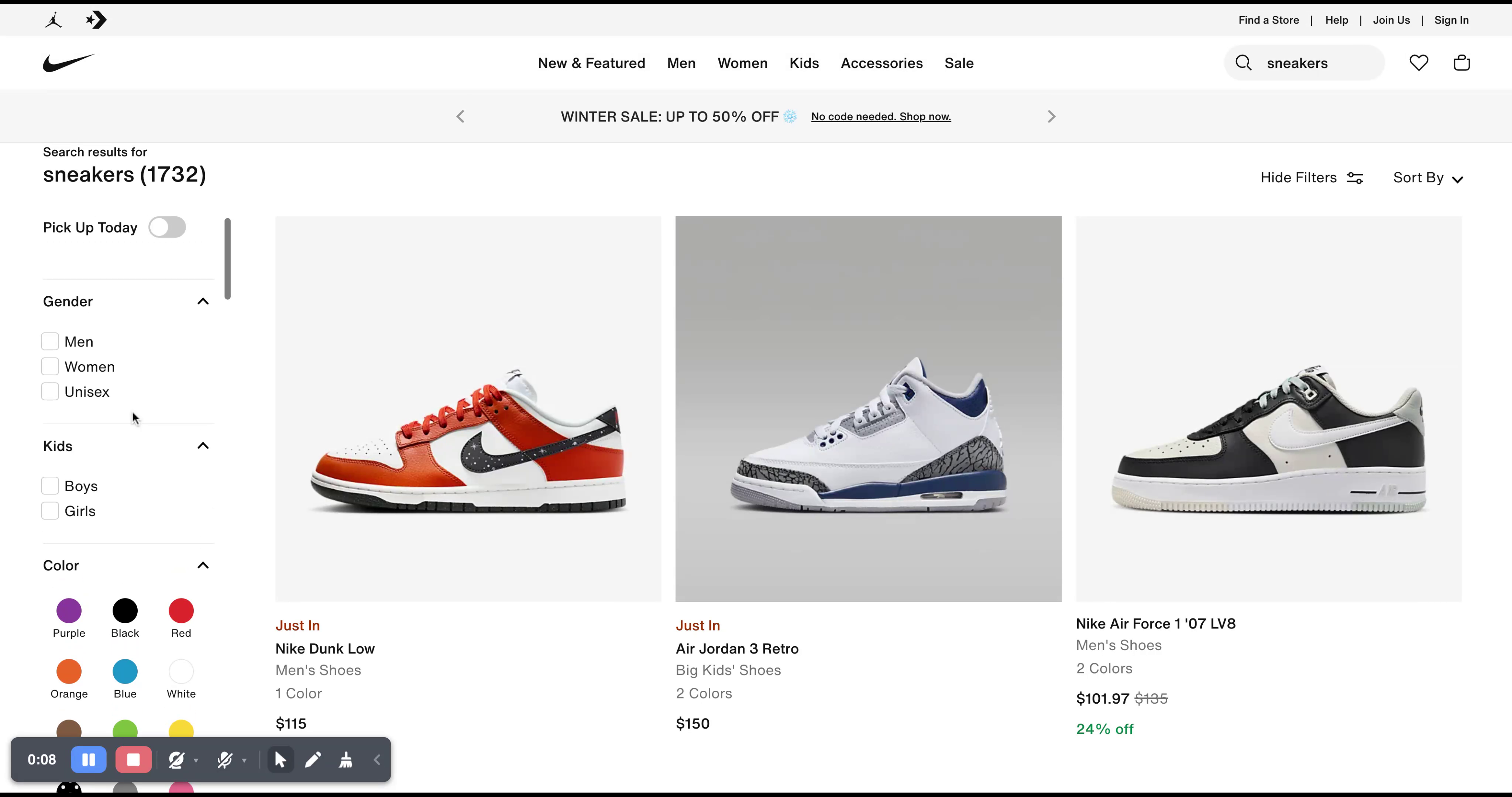}\par
    \includegraphics[width=\linewidth]{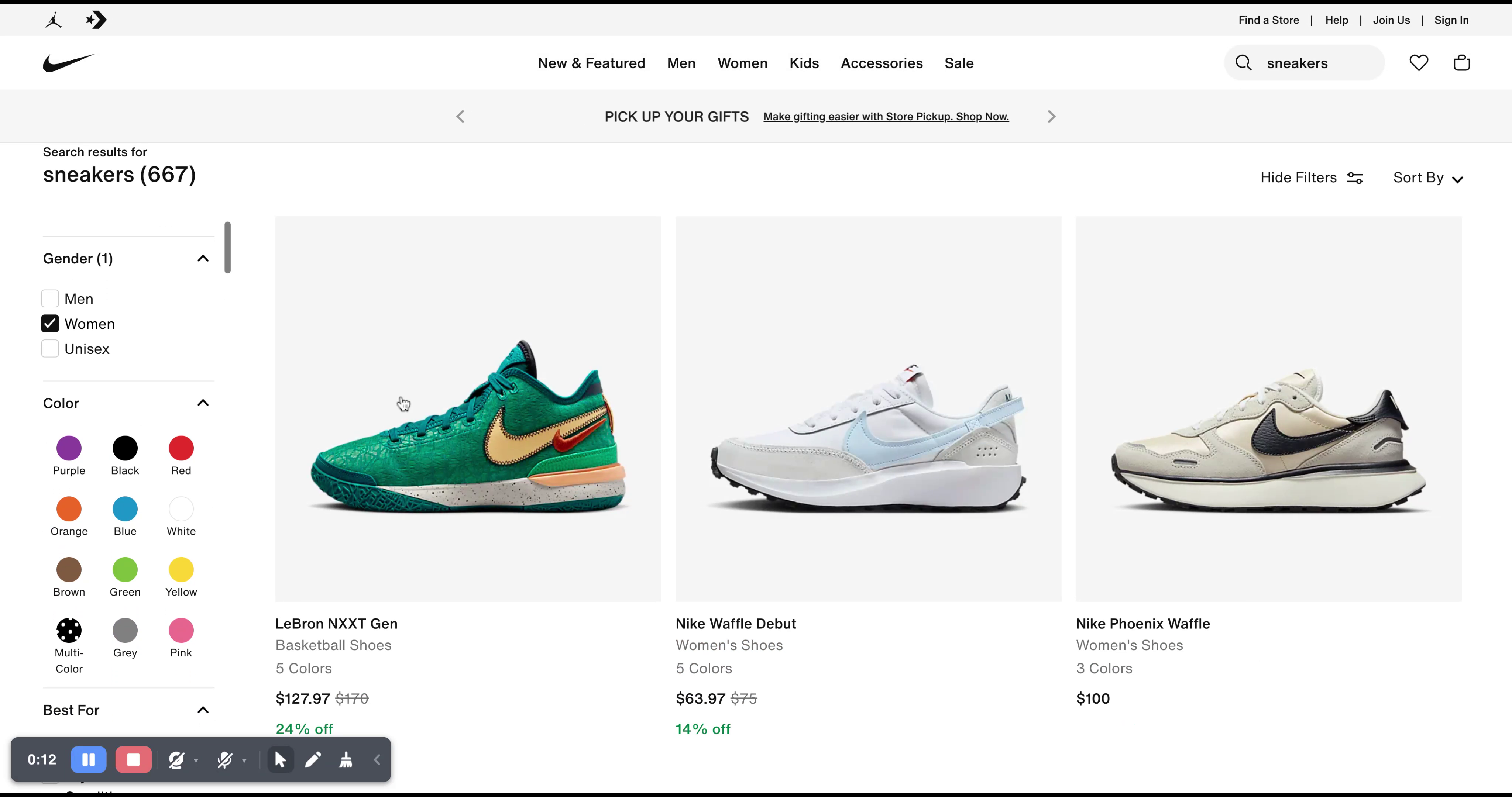}\par
    \includegraphics[width=\linewidth]{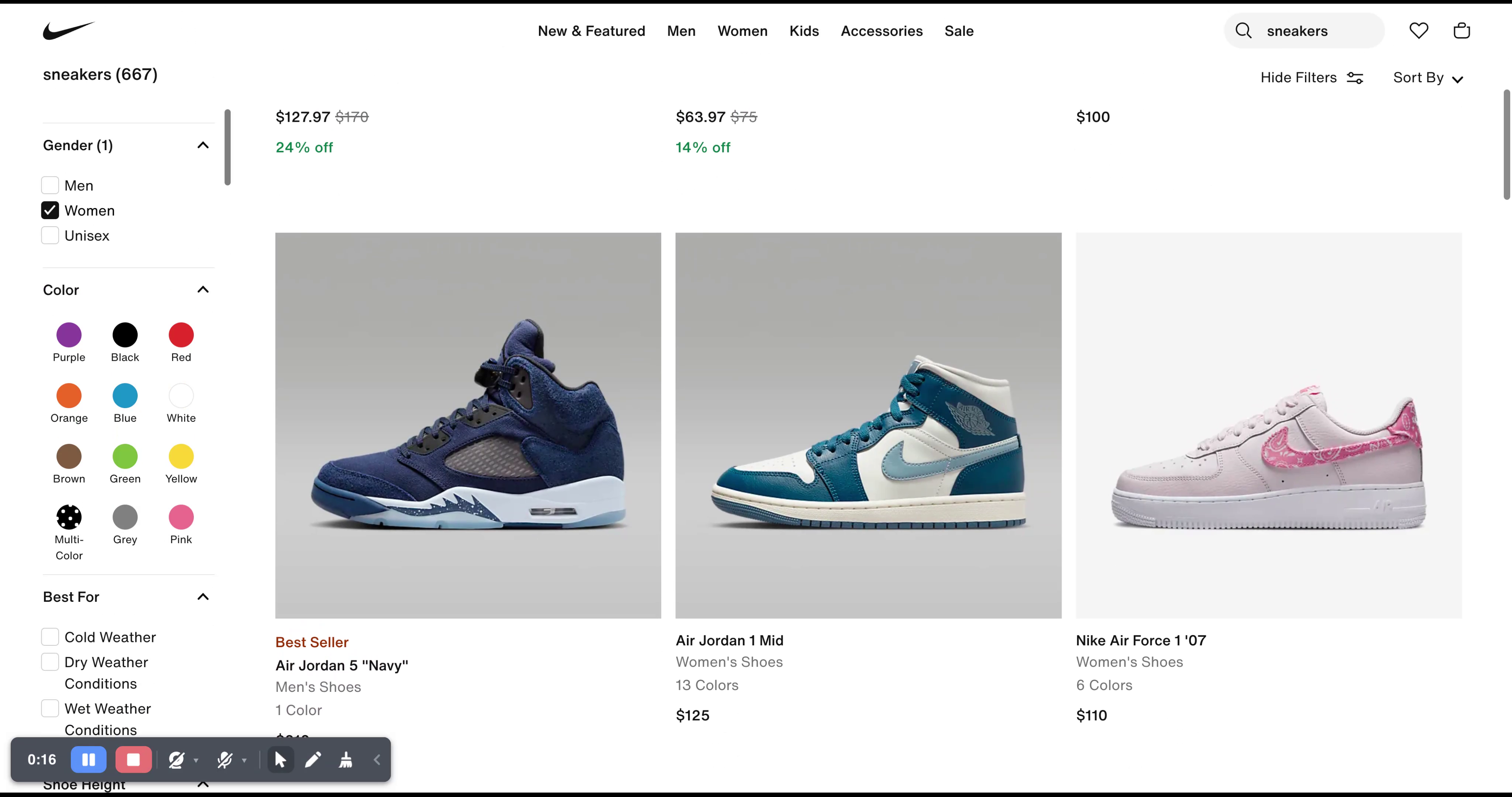}\par
    \includegraphics[width=\linewidth]{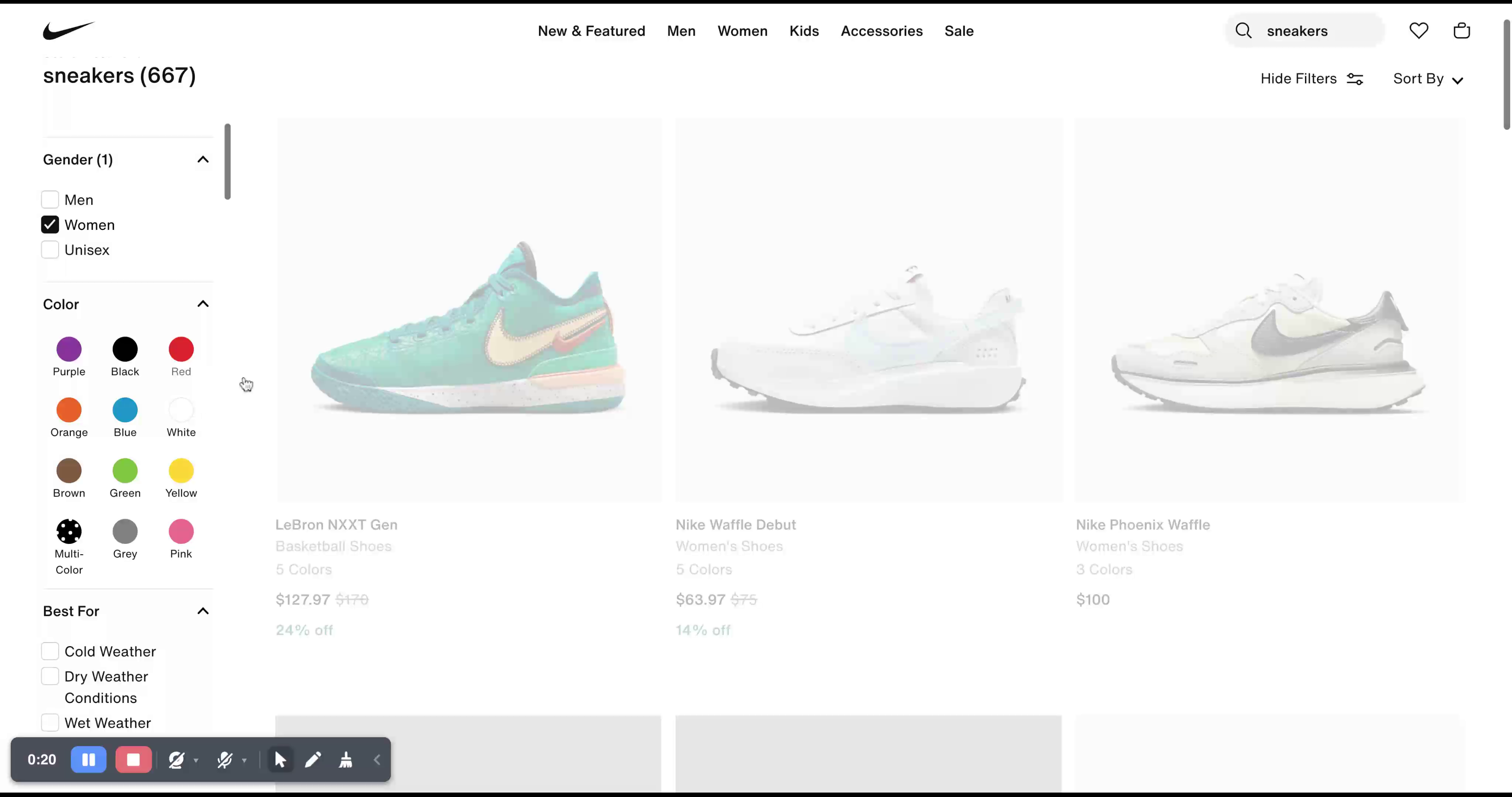}\par
    \includegraphics[width=\linewidth]{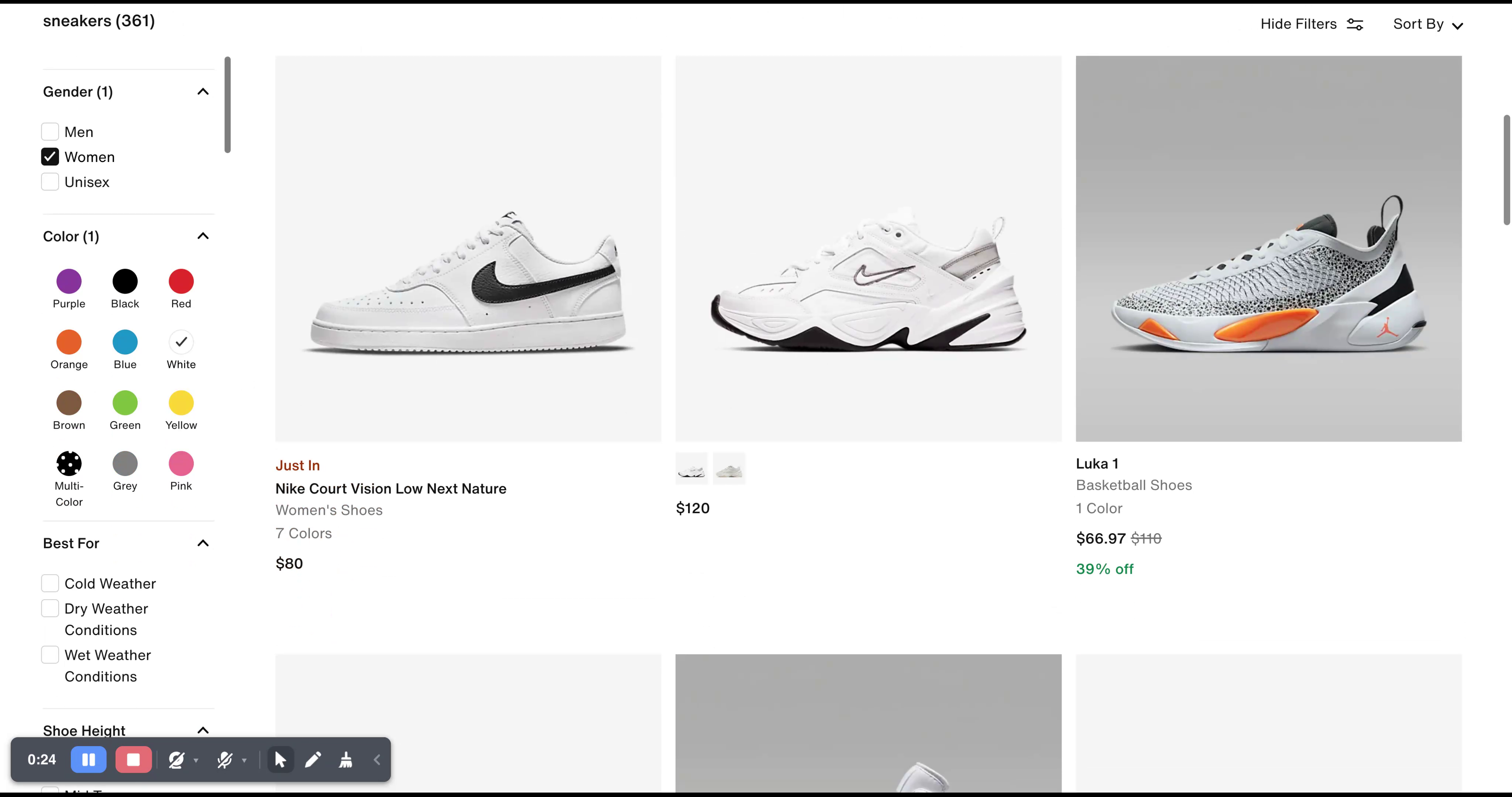}\par
    \includegraphics[width=\linewidth]{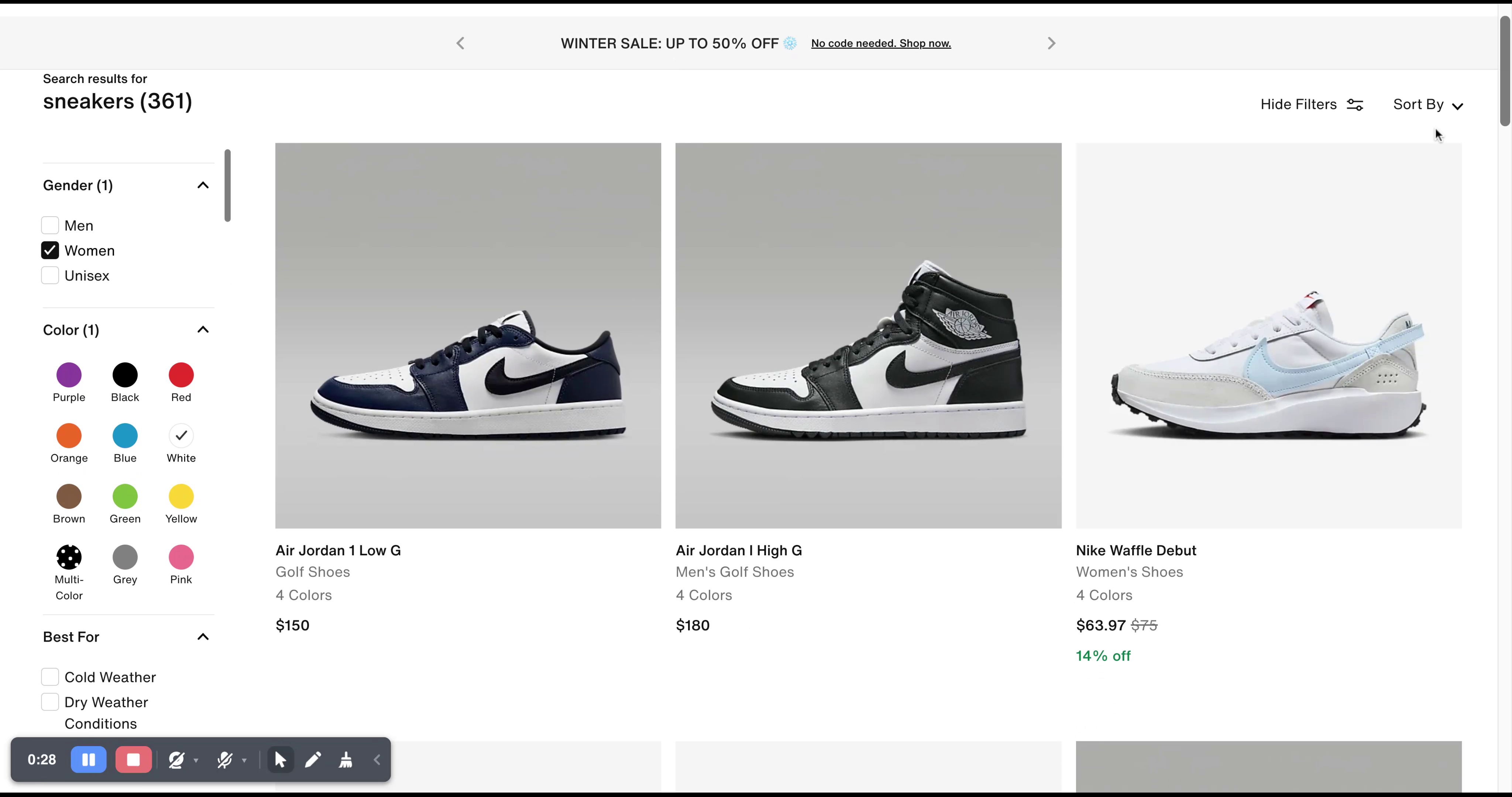}\par
    \includegraphics[width=\linewidth]{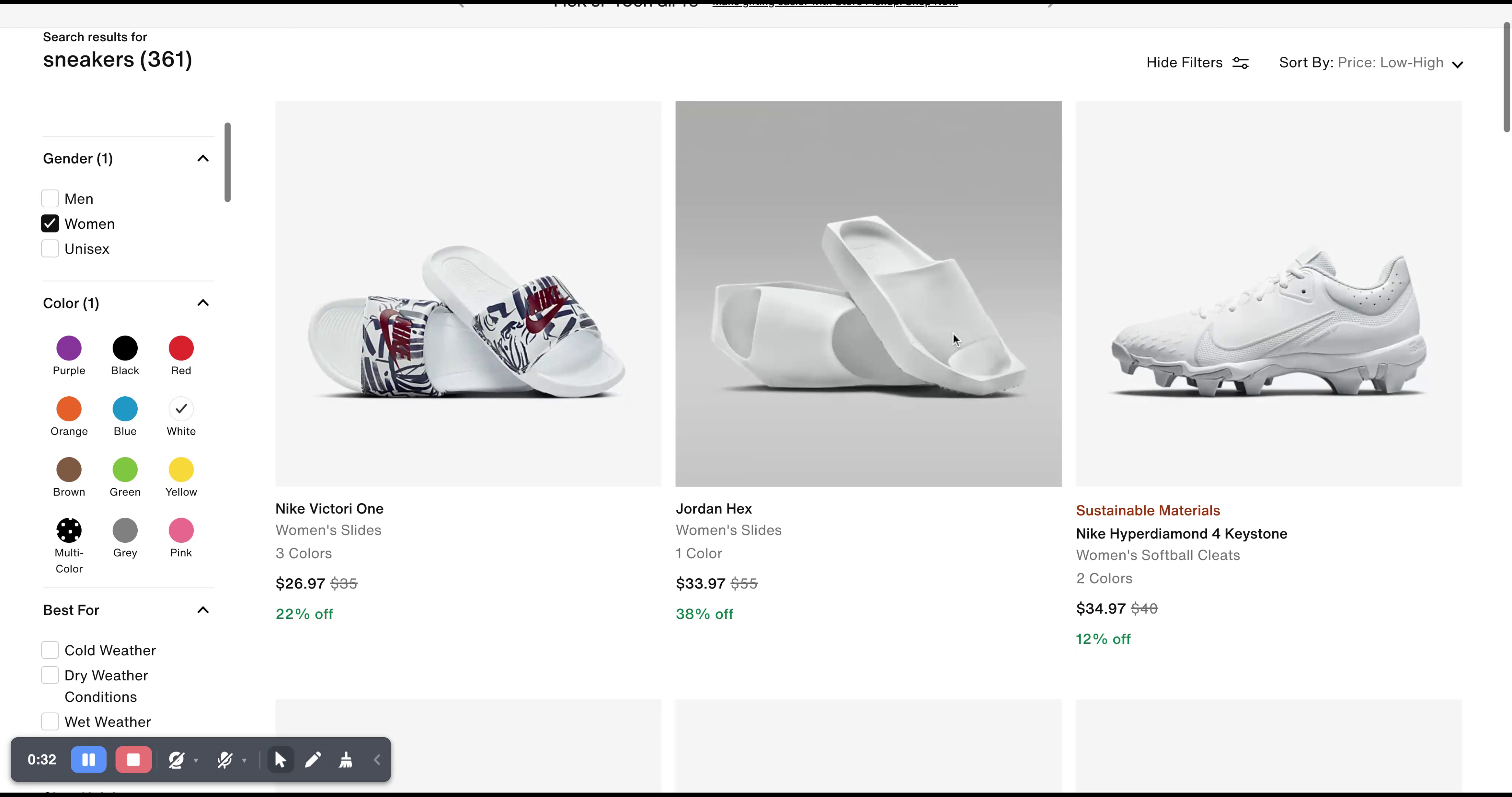}\par
  \end{multicols}
  \caption{Nine screen captures showcasing user interactions with the Nike.com website, each captured at a 4-second interval.}
  \label{fig:3x3_images_multicol}
\end{figure}

\begin{figure}[htbp]
\begin{tcolorbox}[colframe=cyan, colback=white!10]
``json{ \textit{``Product Characteristics"}: ``The products featured are mostly casual and sports footwear, with a focus on style and functionality.",  \colorbox{yellow!50}{\textit{``Lowest Price"}: ``\$63.97}, as seen on the Nike Waffle Debut Women\'s Shoes.",  \colorbox{yellow!50}{\textit{``Highest Price"}: ``\$180}, as seen on the Air Jordan 1 High G Men\'s Golf Shoes.",  \colorbox{yellow!50}{\textit{``Brand Preference"}}: ``The brand showcased is Nike, including its \colorbox{yellow!50}{Air Jordan} line, indicating a preference for this brand.",  \textit{``Product Specifications"}: {    ``Gender``: ``Men, Women, and Unisex options available",    ``Color": ``Multiple color options are available for selection",    ``Best For": ``options like Cold Weather, Dry Weather Conditions, Wet Weather Conditions are available, although specific preferences are not indicated in the images."  },  \textit{``User Reviews and Testimonials"}: ``Not available",  \textit{``Comparisons"}: ``The images showcase different models and prices, suggesting that comparisons can be made within the selection, but specific comparative analysis is not displayed.",  \textit{``Promotions"}: ``Promotions are visible, such as \'WINTER SALE: UP TO 50\% OFF\' and discounts on specific products."}''
\end{tcolorbox}
\caption{Extracted summary of user preferences shaped by activities on the Nike.com website.}
\end{figure}

\begin{figure}[htbp]
  \begin{multicols}{2}
    \includegraphics[width=\linewidth]{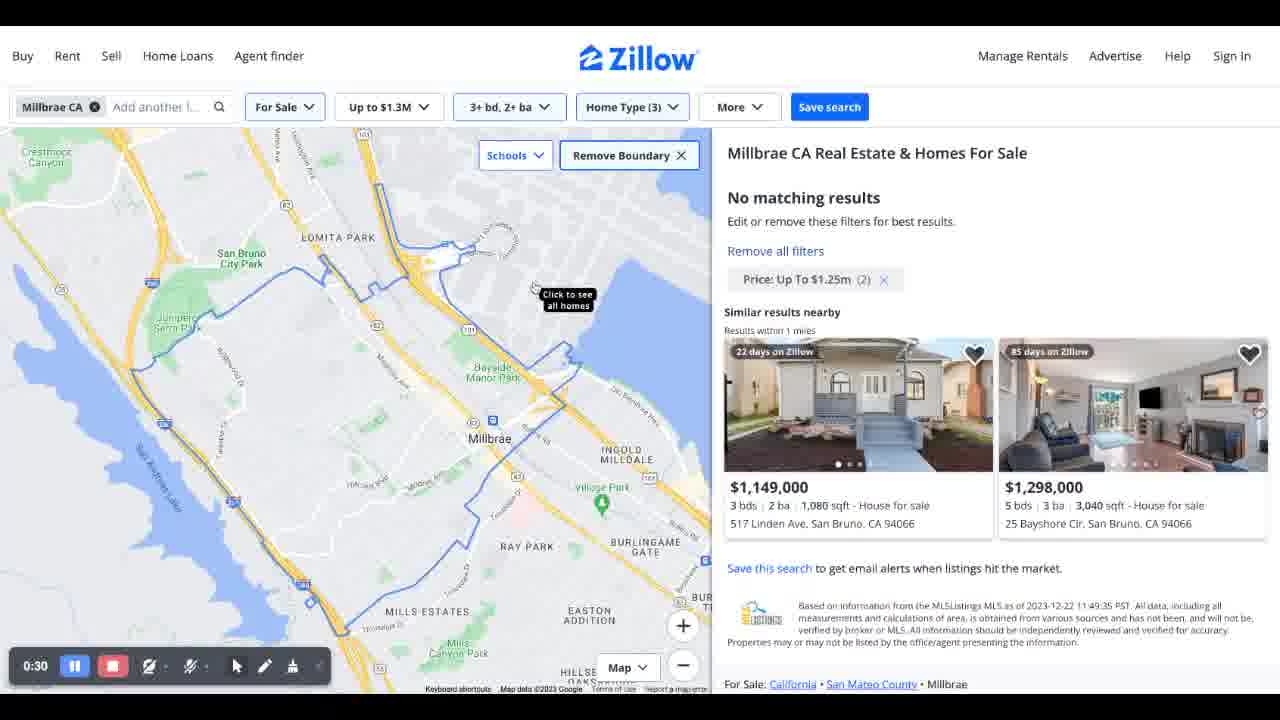}\par
    \includegraphics[width=\linewidth]{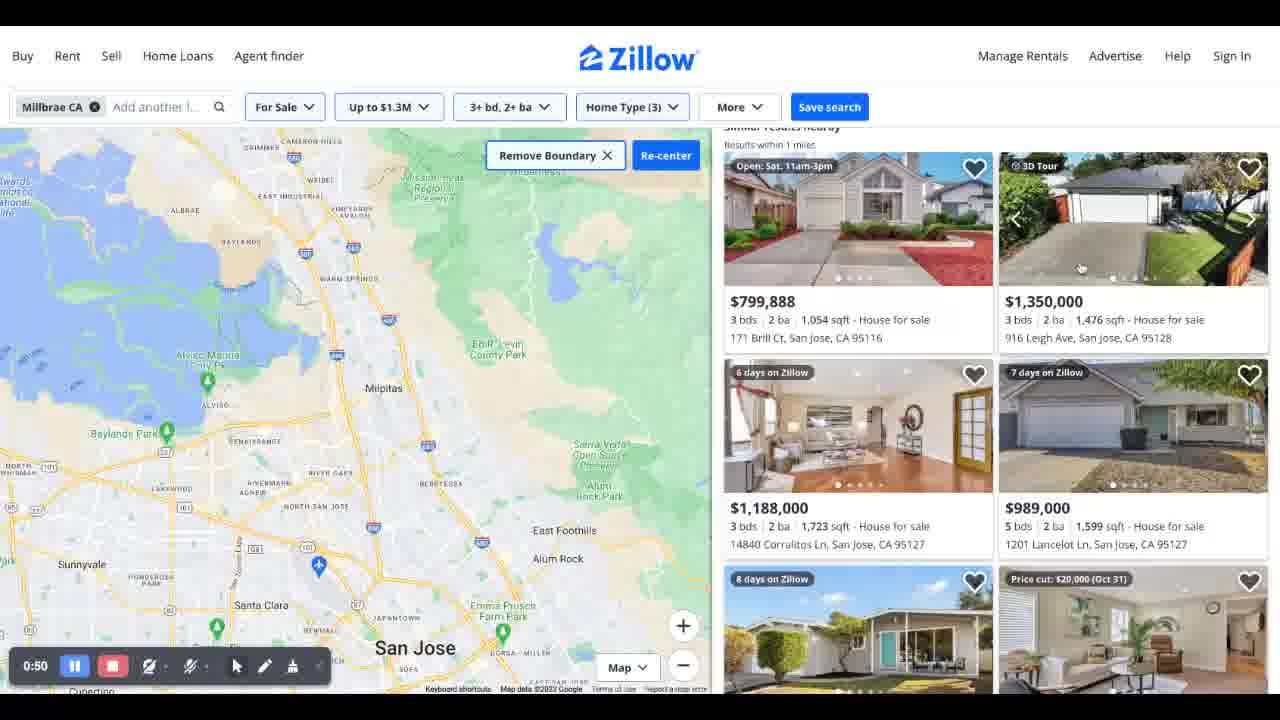}\par
    \includegraphics[width=\linewidth]{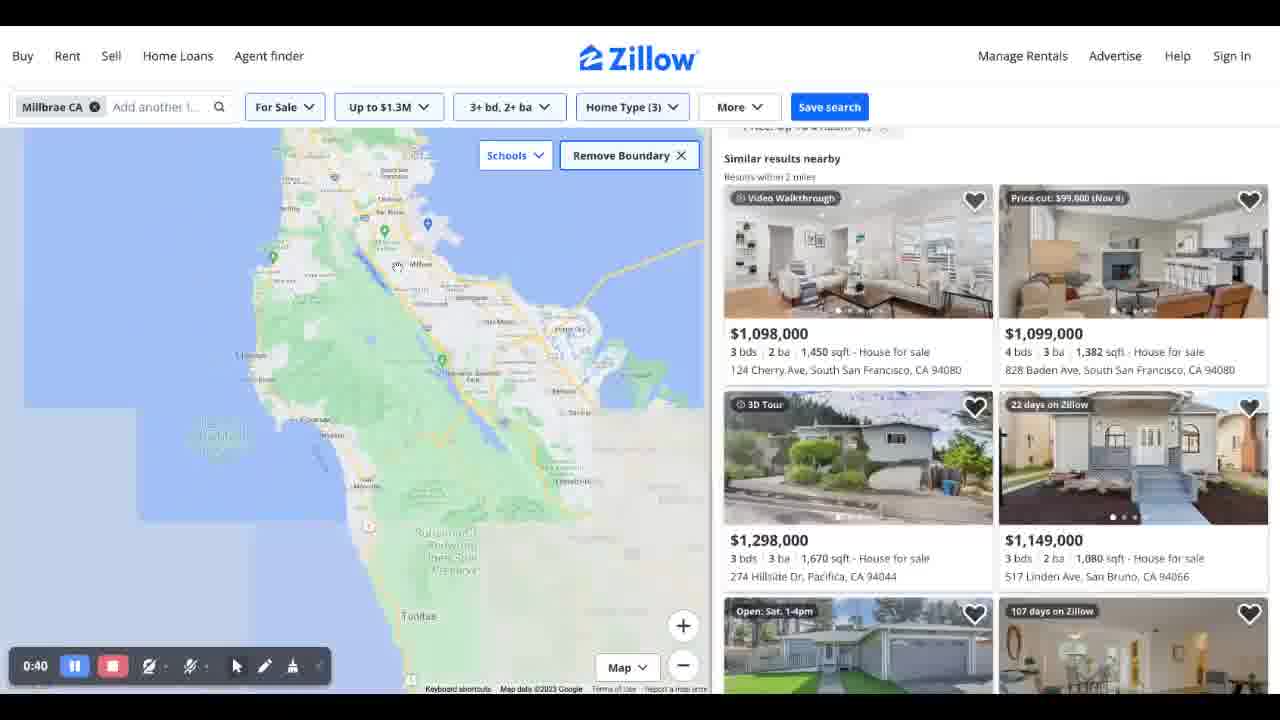}\par
    \includegraphics[width=\linewidth]{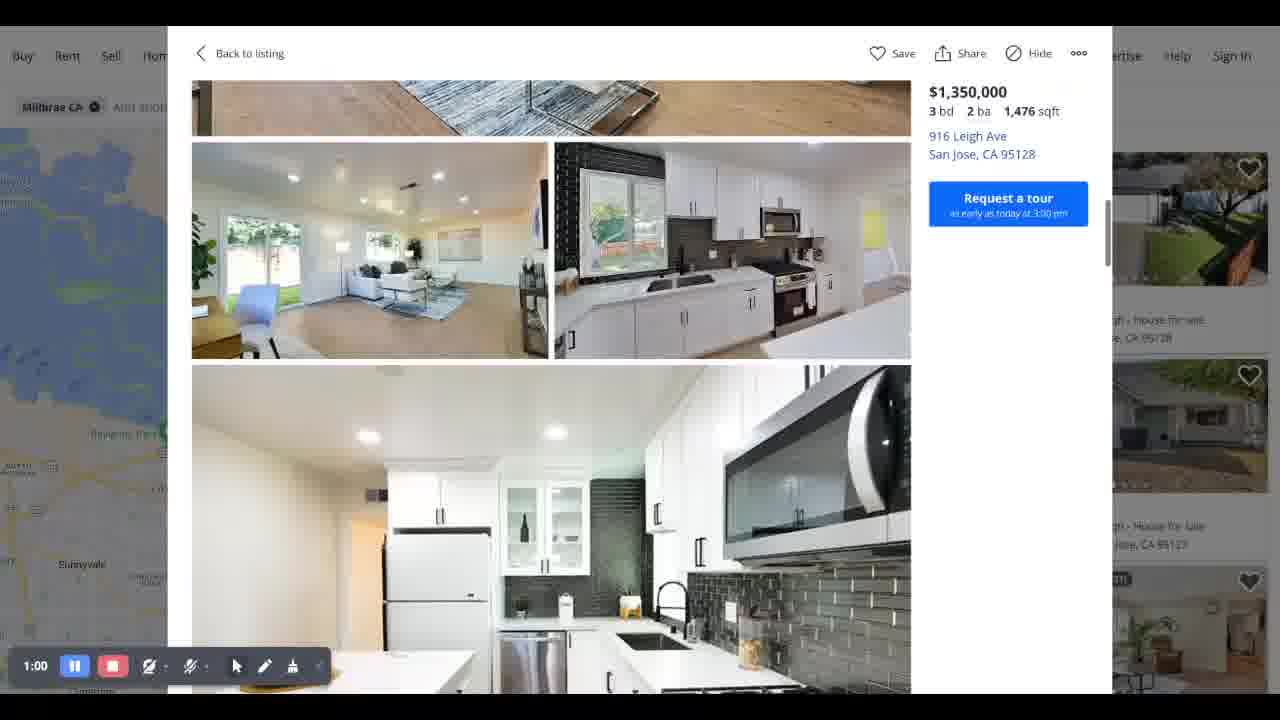}\par
  \end{multicols}
  \caption{Four screen captures showcasing user interactions with the Zillow.com website, each captured at a 10-second interval.}
  \label{fig:2x3_images_multicol}
\end{figure}

\begin{figure}[htbp]
\begin{tcolorbox}[colframe=cyan, colback=white!10]
``Based on the provided images, it appears that the user is searching for real estate properties for sale on a website, likely with certain preferences. The images show a website with real estate listings that include price ranges, the number of bedrooms and bathrooms, square footage, and location. Using the data visible in the images, I will attempt to create JSON-formatted data that represents the user\'s inferred preferences within the limitations of the images.``json{  \colorbox{yellow!50}{\textit{``Product Characteristics"}}: {    ``Location": ``Millbrae, CA; South San Francisco, CA; Pacifica, CA; San Jose, CA",    \colorbox{yellow!50}{``Property Type": ``House",    ``Bedrooms": "3+",    ``Bathrooms": "2+"  }},  \colorbox{yellow!50}{\textit{``Lowest Price"}: “\$799,888",  \textit{``Highest Price"}: ``\$1,350,000"},  \textit{``Brand Preference"}: ``not available",  \textit{``Product Specifications"}: {    ``Square Footage": {      ``Low": ``,054 sqft",      ``High": ``3,040 sqft"    }  },  \textit{``User Reviews and Testimonials"}: ``not available",  \textit{``Comparisons"}: {    ``Available": true,    ``Criteria": ``Price, Location, Size, Bedrooms, Bathrooms"  },  \textit{``Promotions"}: {    ``Price Cut": ``\$20,000 (Oct 31) on a property listed for \$989,000"  }} ‘’Please note that the information above is constructed from the context provided by the images and might lack accuracy in details that are not visible or provide insufficient context. Some categories like \'Brand Preference\' and \'User Reviews and Testimonials\' do not apply in the context of real estate listings and therefore are marked as ``not available." Also, the promotion information is based on limited data indicating a price cut on a specific property, as seen in the images.''
\end{tcolorbox}
\caption{Extracted summary of user preferences shaped by activities on the Zillow.com website.}
\end{figure}

\end{document}